\documentclass{IEEEtran}
\IEEEoverridecommandlockouts
\usepackage{cite}
\usepackage{amsmath,amssymb,amsfonts}
\usepackage{algorithmic}
\usepackage{graphicx,color}
\usepackage{textcomp}
\usepackage{xcolor}
\usepackage{hyperref}
\hypersetup{hidelinks}
\usepackage{algorithm,algorithmic}
\def\BibTeX{{\rm B\kern-.05em{\sc i\kern-.025em b}\kern-.08em
    T\kern-.1667em\lower.7ex\hbox{E}\kern-.125emX}}
\AtBeginDocument{\definecolor{tmlcncolor}{cmyk}{0.93,0.59,0.15,0.02}\definecolor{NavyBlue}{RGB}{0,86,125}}
\usepackage{orcidlink}
\usepackage{siunitx}
\usepackage{pgfplots}
\usepackage{pgfplotstable}
\pgfplotsset{compat=newest}
\usepgfplotslibrary{groupplots}

\renewcommand{\thesubsection}{\thesection-\Alph{subsection}}

\begin{document}

\markboth{}{Heraldo C. A. Costa {et al.}}

\title{Modeling Micro-Doppler Signature of Multi-Propeller Drones in Distributed ISAC}

\author{\IEEEauthorblockN{
Heraldo C. A. Costa\IEEEauthorrefmark{1}\orcidlink{0009-0002-6186-5780}, Student Member, IEEE,
Saw J. Myint\IEEEauthorrefmark{1}\orcidlink{0009-0007-3788-7126}, Student Member, IEEE,
Carsten Andrich\IEEEauthorrefmark{1}\orcidlink{0000-0002-4795-3517},
Sebastian W. Giehl\IEEEauthorrefmark{1}\orcidlink{0009-0008-1672-1351},
Maximilian Engelhardt\IEEEauthorrefmark{1}\orcidlink{0009-0002-9440-8615},
Christian Schneider\IEEEauthorrefmark{1}\orcidlink{0000-0003-1833-4562},
Reiner S. Thom\"a\IEEEauthorrefmark{1}\orcidlink{0000-0002-9254-814X}, Life Fellow, IEEE
}
\IEEEauthorblockA{\IEEEauthorrefmark{1}
Institute for Information Technology and Thuringian Center of Innovation in Mobility,\\
Technische Universität Ilmenau, Ilmenau, Germany\\
\scriptsize{\hfill \\This research is funded by the BMBF project 6G-ICAS4Mobility, Project No.16KISK241, by the Federal State of Thuringia, Germany,\\and by the European Social Fund (ESF) under grants 2017 FGI 0007 (project ``BiRa'') and 2021 FGI 0007 (project ``Kreat\"or'').}}
}

\IEEEoverridecommandlockouts
\IEEEpubid{
\makebox[\textwidth]{\raisebox{-2ex} {\scriptsize 979-8-3503-2920-9/24/\$31.00~\copyright2025 IEEE | DOI: 10.1109/JSTEAP.2025.3604407} \hfill}}

\maketitle

\begin{abstract}
Integrated Sensing and Communication (ISAC) will be one key feature of future 6G networks, enabling simultaneous communication and radar sensing. 
The radar sensing geometry of ISAC will be multistatic since that corresponds to the common distributed structure of a mobile communication network. 
Within this framework, micro-Doppler analysis plays a vital role in classifying targets based on their micromotions, such as rotating propellers, vibration, or moving limbs. However, research on bistatic micro-Doppler effects, particularly in ISAC systems utilizing OFDM waveforms, remains limited. Existing methods, including electromagnetic simulations, often lack scalability for generating the large datasets required to train machine learning algorithms. 
To address this gap, this work introduces an OFDM-based bistatic micro-Doppler model for multi-propeller drones. 
The proposed model adapts the classic thin-wire model to include bistatic sensing configuration with an OFDM-like signal. Then, it extends further by incorporating multiple propellers and integrating the reflectivity of the drone’s static parts. 
Measurements were performed to collect ground truth data for verification of the proposed model. 
Validation results show that the model generates micro-Doppler signatures closely resembling those obtained from measurements, demonstrating its potential as a tool for data generation. In addition, it offers a comprehensive approach to analyzing bistatic micro-Doppler effects.
\end{abstract}



\maketitle

%
%

\section{INTRODUCTION}\label{section:Introduction}
\IEEEPARstart{I}{ntegrated} Sensing and Communication (ISAC), also abbreviated as ICAS, is considered one of the key features of future 6G mobile communication \cite{FanLiu2023ICASbook}, \cite{Thoma2}. 
In essence, ISAC is a means of radar detection and localization of passive objects (“targets”) that are not equipped with a radio tag. 
The radio signals transmitted by the mobile radio nodes for communication purposes are reused for target illumination. 
At the same time, the mobile communication network ensures data fusion and processing. 
ISAC has the potential to create comprehensive situational awareness, increased efficiency, and safety in public transportation, such as road traffic scenarios or lower altitude air space (U-space) monitoring by complementing upcoming Uncrewed Aircraft Systems (UAS) traffic management systems. 
Moreover, there are many other promising use cases of ISAC for public security and safety applications and for increased efficiency in mobile radio access \cite{Shatov2024}.

ISAC takes advantage of the radio access schemes of multiuser mobile radio. 
This way, a distributed Multi-Sensor Integrated Sensing and Communication (MS ISAC) network is established. 
According to the common distribution of user equipment and infrastructure in mobile communications, ISAC sensing geometry is bistatic and multistatic. 
Monostatic sensing is also discussed but is likely to become the exception, as full-duplex access has not yet been introduced in mobile communications and will be problematic with the Orthogonal Frequency-Division Multiplexing (OFDM) waveforms used there. 
Therefore, in the most general case of MS ISAC \cite{thoma2023Dis}, it becomes a distributed Multiple Input Multiple Output (MIMO) radar \cite{Haimovich2008}, which enables the estimation of the full three-dimensional (3D) dynamic target state vector. 
A related advantage is target-related diversity gain, which increases detection probability and supports target recognition. 
The inherent bistatic nature of ISAC somehow resembles the well-known passive radar principle. 
A related, very efficient ISAC implementation was first proposed as Cooperative Passive Coherent Localization (CPCL) in \cite{thoma2019cooperative}.
Target recognition will be based on specific target-related features that are related to spatial and delayed distribution of the target reflectivity. 
While distributed sensors, wide bandwidth, and, perhaps, polarimetry recognize the static target shape, the time-variant target response resulting from inherent time variability of targets such as drones, cars, or walking persons requires wideband real-time measurement. 
Time variability is also identified as micro Doppler \cite{Chan2019}. 
Micro-Doppler refers to the small fluctuations in the Doppler shift caused by the target’s local moving parts, e.g., the rotating blades of drones or the flapping wings of birds. 
Those fluctuations provide valuable information to be used for target classification  \cite{Liu2024},  \cite{xsmith2008radar}

Despite the obvious advantage of bistatic sensing geometry for target identification, there is currently only a limited amount of data and research on bistatic micro-Doppler compared to monostatic micro-Doppler. 
Besides bistatic real-time measurement systems such as \cite{andrich2024bira}, corresponding modeling and simulation tools are necessary for the training of classification algorithms, such as machine learning and deep learning layouts, that require large quantities of labeled training data \cite{de2022next}.
\IEEEpubidadjcol
An important case in micro-Doppler modeling is the study of the micro-Doppler of targets with rotating propellers, such as helicopters, jet aircraft, wind turbines, and drones.
Some works have used electromagnetic simulations based on Computer-Aided Design (CAD) models to generate synthetic data of rotating propellers. 
\cite{Pouliguen2002} describes an application of Physical Optics (PO) and the Method of Equivalent Currents (MEC) to calculate the time-varying Radar Cross-Section (RCS) of a helicopter rotor.
\cite{karabayir2016micro} combines PO simulation from CAD models in different positions to generate the time-varying reflectivity of a propelled aircraft. 
\cite{schroder2016numerical} utilizes the Method of Moments (MoM) as well as the Physical Optics method to simulate the micro-Doppler of a multi-propeller drone with 512 discrete propeller positions.
Similarly, \cite{Speirs2018} uses a hybrid Finite Element Method (FEM) and MoM scheme, and also considers 512 different uniformly spaced propeller positions through 360°, and then compares micro-Doppler returns from different drone models.
\cite{Petrovic2021} employs multiple full-wave Electromagnetic (EM) simulations of a multi-propeller drone to create the returns from half of the full rotation of all propellers. 
These methods have the advantage of taking into account the complex structures of the objects and reproducing effects such as diffraction and shadowing. 
However, they are extremely time-consuming and hard to implement. As a consequence, using these methods to generate a large amount of data to train machine learning algorithms becomes unfeasible.

Therefore, many authors have presented or used simplified methods to generate synthetic data on rotating propellers. 
In \cite{chen2000time, chen2006micro, White2022, Wei2024}, the authors have analyzed the modeling of each rotating propeller as a single point target.
\cite{Lehmann2020} and \cite{Plotnitskaya2021} have modeled propellers as a finite set of point scatterers, defining the distance between adjacent points as a pre-defined fraction of the wavelength.
\cite{moore2022pointclouds} has presented a simulation approach that models all components of the UAS, including its static parts, as a finite set of point scatterers derived from 3D CAD models.

However, the most used method for modeling rotating propellers is the thin wire model, which considers a propeller as a homogeneous, linear, rigid antenna, that can be replaced by an infinite set of point scatterers.
This approach was used in \cite{xsmith2008radar, White2022, Schneider1988, martin1990analysis, misiurewicz1997analysis, Jianjiang2005, Costa2008, Chen2012, Wang2015, Cai2019, Ummenhofer2019, Point2017}.

From all of these works, only the models presented in \cite{Pouliguen2002} and \cite{Plotnitskaya2021} cover bistatic geometry, while \cite{Clemente2012} uses bistatic geometry but do not let clear which are the assumptions considered in their modeling.
Only \cite{Wei2024} discusses ISAC and OFDM, but only in monostatic geometry.
The literature still misses works that study the problem of micro-Doppler modeling in the context of ISAC, including both OFDM waveforms and bistatic geometry.

The present work proposes an OFDM bistatic micro-Doppler model for multi-propeller drones, developed from the classic monotonic thin wire model for monostatic radars, intended to become an interesting tool for future development of ISAC target classification algorithms. 
We proposed the model of a single propeller in \cite{Heraldo2024modelling}, that assumes each propeller blade behaving as a thin-wire scatterer with negligible thickness and uniform material properties.
This provides a computationally efficient approximation, though real-world blades exhibit material-dependent scattering effects that are not captured.
In the current paper, we extend this model by including multiple propellers and two different ways of generating the static reflectivity of the drone body.

\autoref{section:Model} introduces the mathematical description of the OFDM bistatic micro-Doppler model of a rotating propeller.
In \autoref{section:Drone_Model}, we propose a model that expands the propeller formulation to create the multi-propeller drone micro-Doppler, with validation against measurements in \autoref{section:validation}.
\autoref{section:AdvantagesDisadvantages} presents the advantages and disadvantages of the proposed model.
Conclusion and outlook can be found in \autoref{section:Conclusion}.

\section{MICRO-DOPPLER MODEL OF A SINGLE ROTATING PROPELLER}\label{section:Model}
\subsection{MONOSTATIC MONOTONIC THIN WIRE PROPELLER MODEL}\label{subsection:classic_model}

    Suppose that a radar transmitter (Tx) produces a signal described as:
            
    \begin{equation}\label{eq:base_radar_signal}
        x(t) = \gamma_0 \mathrm{e}^{\mathrm{j}\omega_0t},
    \end{equation}
    where $\gamma_0$ is the amplitude of the transmitted signal, $\omega_0 = 2\pi f_0 = 2\pi  \frac{c}{\lambda_0}$, $f_0$ is the transmitter central frequency, $\lambda_0$ is the wavelength, and $c$ is the light speed.
        
    If a scatterer point target P has a range $R_P^{(m)}(t)$ with respect to a radar, where $^{(m)}$ stands for monostatic, its backscattered returns to the radar receiver (Rx) can be expressed as \cite{Richards2014} 
    
    \begin{equation}\label{eq:sP}
        y_P(t) = \gamma_P(t) \mathrm{e}^{\mathrm{j}\omega_0 \left[t-\frac{2 R_P^{(m)}(t)}{c}\right]} + z(t),
    \end{equation}
    where $z(t)$ is the noise, and $\gamma_P(t)$ is the amplitude of the returned signal, given by \cite{braun2014ofdm}
    
    \begin{equation}\label{eq:radar_equation}
        \gamma_P(t) = \gamma_0 \sqrt{\frac{c \ \sigma_P}{(4\pi^3)R_P(t)^4 f_0^2}},
    \end{equation}
    $\sigma_P$ is the RCS of the point scatterer $P$.

    From now on, the model that will be described assumes that the target is an infinite set of multiple independent point scatterers.
    For this assumption to be valid, the wavelength must be smaller than the target.
    
    As the interaction among different scattering centers is very weak, the scattering can be assumed to be a local linear process, i.e., the target returns in a given range resolution cell is the superposition (coherent sum) of each independent scattering center in that resolution cell at that instant \cite{Jianjiang2005}.

    In this case, the target’s total scattering echo is given by the coherent sum of all independent scattering centers at that instant. 
    Therefore, considering a blade $i$ as the single rigid, homogeneous, linear antenna, thus a line of infinite point scatterer going from the center \textbf{O} until a point \textbf{P\textsubscript{tip}}, distant $l$ from the center \textbf{O}. 
    The mathematical model for the monostatic slow-time returns (micro-Doppler signature) of a rotor with $N_B$ rotating blades of length $L_B$ each is derived in \cite{Schneider1988}: 
    
    \begin{gather}\label{eq:SN}
        \tilde{h}_i(t) = \sum_{i=1}^{N_B} \bigg\{ \tilde{\gamma}_i(t) \mathrm{e}^{\mathrm{j}\frac{\omega_0}{c}\left(-2R_O+L_B \cos(\omega t+\varphi_B (i)) \cos(\psi)\right)} \nonumber \\
        \cdot L_B \operatorname{sinc}\left[\frac{\omega_0 L_B}{c} \cos(\omega t+\varphi_B(i)) \cos(\psi)\right] \bigg\} + \tilde{z}(t),
    \end{gather}  
    where $R_O$ is the range $R_P^{(m)}(t)$ of the rotation center \textbf{O}, $\omega$ is the angular velocity of the blades, $\psi$ is the elevation angle from the radar to the rotation center with respect to the rotation plane, and $\tilde{\gamma}_i(t)$ and $\tilde{z}(t)$ are $\gamma_P(t)$ and $z(t)$ after matched filtering, and the initial azimuth angle of each propeller $\varphi_B(i)$ is given by
     \begin{equation} \label{eq:phi_B_i}
        \varphi_B(i) = \varphi_0 + 2\pi\frac{i-1}{N_B},
    \end{equation} 
    where $\varphi_0$ is the initial angular position of the propeller.

    As the phase of the received signal depends on the total time spent to propagate through the path Tx-Target-Rx, changes in $R_P^{(m)}(t)$ generate changes in the phase of the signal. If the target is moving, the range $R_P^{(m)}(t)$ corresponding to each symbol echo is changing. As a result of this, the phase term becomes a function of $R_P^{(m)}(t)$, and the Doppler frequency is given by the derivative of this phase with respect to the slow-time \cite{Qu2019}.

    Therefore, this model represents the monostatic slow-time response of a rotor with $N_B$ blades when our radar/sensor uses a single-tone waveform.  

    Since the range resolution in the direction of the bistatic bisector is given by $\Delta R = c/[2B\cos{(\beta/2)}]$\cite{willis2005bistatic} when a low signal bandwidth $B$ is used, the range resolution is finer than the target dimensions and therefore, the target returns in a given slow-time instant are entirely contained in a single resolution cell.
    In coherent processing, consecutive returns are concatenated along the slow-time dimension, forming a vector, and the target micro-Doppler signature can be analyzed unidimensionally.
    The model presented in this section can be used with a good level of success in these narrowband cases \cite{Point2017}.
    
    On the other hand, a broadband scenario leads to an extended response in the fast-time/range dimension, which stands for the High-Range Resolution (HRR) case.   
    Different from the classic thin wire model, the model proposed in the next sections can be used to generate HRR data.

\subsection{BISTATIC GEOMETRY OF A ROTATING POINT SCATTERER}\label{subsection:geometry}

    To derive a similar model for the broadband bistatic case, it is first necessary to have a geometrical model for the bistatic range of a single rotating point.
    
    \begin{figure}[t!]
        \centering
        \includegraphics[width=8cm]{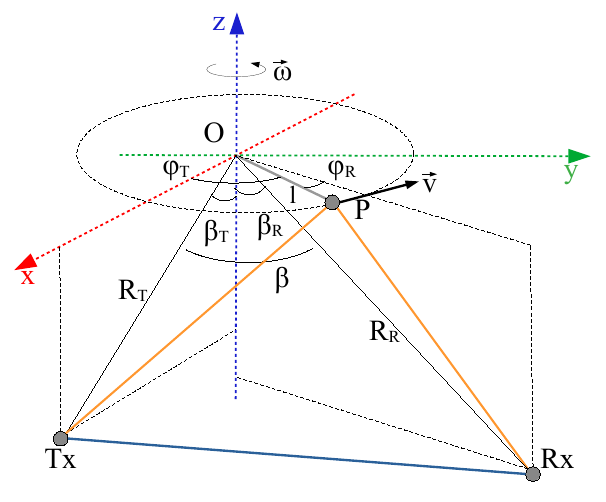}
        \caption{Bistatic rotating point geometry}
        \label{fig:rot_pt1}
    \end{figure}
    
    Such a math model was presented in \cite{Ai2011}, where the author develops the mathematical formulation of the bistatic range of a point scatterer \textbf{P}, in a bistatic radar environment, rotating around a point \textbf{O}, according to \autoref{fig:rot_pt1}. This bistatic range is
    \begin{equation} \label{eq:R_t}
        R_P^{(b)}(t) \approx R_O - A_B l \cos(\omega t+\varphi_B),
    \end{equation}
    where $A_B = \sqrt{4\cos(\beta/2)^2-(\cos\beta_T + \cos\beta_R)^2}$, $R_O = R_T + R_R$ is the total bistatic range of the rotation center \textbf{O}, and
     \begin{equation} \label{eq:phi_B}
        \varphi_B = \varphi_T - \arctan\left(\frac{\sin\beta_R \sin(\varphi_T - \varphi_R)}{\sin\beta_T + \sin\beta_R \cos(\varphi_T - \varphi_R)}\right) .
    \end{equation}

    Two important things to note are that, in this formulation, there is no restriction about the relative position between the rotation plane and the Tx / Rx positions, and that $R_O$ is also time-dependent if the target is moving.    
    
    Based on \eqref{eq:R_t}, and using the same method as in \autoref{subsection:classic_model} for the monostatic case, it is now possible to derive a model for bistatic returns of a rotor with $N_B$ rotating blades of length $L_B$ each.

\subsection{HIGH RESOLUTION BISTATIC RETURNS}\label{subsection:high_resolution_bistatic}

    In this case, the returns from the target are divided in more than one range bin. Therefore, the scattering is not only a function of slow-time, but also a function of delay (fast-time).

    The fast-time delay $\tau_P$ related to the point scatterer $P$ in \autoref{fig:rot_pt1}, due to its bistatic range, is given by
    
    \begin{equation}\label{eq:tauP}
        \tau_P = \frac{R_P^{(b)}(t)}{c}.
    \end{equation}

    Applying \eqref{eq:R_t}, and substituting $\approx$ by $=$ for simplicity, we have
    
    \begin{equation}\label{eq:tauP1}
        \tau_P = \frac{R_O - A_B.l\cos(\omega t+\varphi_B)}{c}.
    \end{equation}
    The effect of ignoring the inequality is analyzed in \mbox{\autoref{subsection:Inequality}}.

    If the range resolution is given by $\Delta R$, and the fast-time delay corresponding to this range resolution is given by $\Delta\tau = \frac{\Delta R}{c}$, then the return value sampled by the radar at the delay $\tau$ is given by the coherent sum of the returns of all the scatterer points whose fast-time delay is contained in the interval $\tau - \frac{\Delta\tau}{2} < \tau_P < \tau + \frac{\Delta\tau}{2}$.

    Applying this concept in (\ref{eq:sP}), we have that the return contribution of the scatter point P in the delay $\tau$ is given by:            

    \begin{equation}\label{eq:S_HRR1}
        y_P(t,\tau) = \gamma_P \exp\left[j\omega_0 \left(t-\tau_P\right)\right]\operatorname{rect}(\frac{\tau-\tau_P}{\Delta\tau}) + z(t)
    \end{equation}
    
    Applying (\ref{eq:tauP1}), we have
    \begin{align}\label{eq:S_HRR2}
        y_P(t,l,\tau) = \gamma_P \mathrm{e}^{\mathrm{j}\omega_0\left(t - \frac{R_O - A_B.l\cos(\omega t+\varphi_B)}{c}\right)}\nonumber\\
        \operatorname{rect} \left(\frac{\tau - \frac{R_O - A_B.l\cos(\omega t+\varphi_B)}{c}}{\Delta\tau}\right) + z(t).
    \end{align}
    
    Once again, considering a blade $i$ as an unidimensional line of infinite point scatters, we can calculate its returns by integrating the returns of all point scatters over the length of the blade:
    
    \begin{align} \label{eq:S_HRR3}                  
            y_i(t,\tau) = \gamma_i \int_{0}^{L_B} \mathrm{e}^{\mathrm{j}\omega_0\left(t - \frac{R_O - A_B.l\cos(\omega t+\varphi_B)}{c}\right)}\nonumber\\
            \operatorname{rect} \left(\frac{\tau - \frac{R_O - A_B.l\cos(\omega t+\varphi_B)}{c}}{\Delta\tau}\right) dl + z(t).
    \end{align}

    As, from (\ref{eq:tauP1}), we have
    \begin{equation}\label{eq:tauP2}
        l = \frac{R_O - \tau_P.c}{A_B\cos(\omega t+\varphi_B)},
    \end{equation}
    then, if $l_1$ and $l_2$ correspond to the limits of the blade length comprised by the pulse length in the delay $\tau$, we can rewrite (\ref{eq:S_HRR3}) as    
    \begin{equation} \label{eq:S_HRR4}                  
            y_i(t,\tau) = \gamma_i \int_{l_1}^{l_2} \mathrm{e}^{\mathrm{j}\omega_0\left(t - \frac{R_O - A_B.l\cos(\omega t+\varphi_B)}{c}\right)} dl + z(t),
    \end{equation}
    where, analyzing individually all possible cases, the lower limit of the integral $l_1$ will always be the median of the set    
    \begin{equation}\label{eq:l1}
        \bigg\{0, \frac{R_O - (\tau-\frac{\Delta\tau}{2}).c}{A_B\cos(\omega t+\varphi_B)}, \frac{R_O - (\tau+\frac{\Delta\tau}{2}).c}{A_B\cos(\omega t+\varphi_B)}\bigg\},
    \end{equation}
    and the upper limit $l_2$ will always be median of the set    
    \begin{equation}\label{eq:l2}
        \bigg\{L_B, \frac{R_O - (\tau-\frac{\Delta\tau}{2}).c}{A_B\cos(\omega t+\varphi_B)}, \frac{R_O - (\tau+\frac{\Delta\tau}{2}).c}{A_B\cos(\omega t+\varphi_B)}\bigg\}.
    \end{equation}

    We use the median here as a way of finding the intermediate value among the three options available.
    Proceeding with the solution, we obtain            
    
    \begin{align} \label{eq:S_HRR5} 
        y_i&(t,\tau) = \gamma_i \mathrm{e}^{\mathrm{j}\omega_0\left(t - \frac{R_O}{c}\right)}\nonumber\\&
        \int_{l_1}^{l_2} \mathrm{e}^{\mathrm{j}\frac{\omega_0}{c}A_Bl.\cos(\omega t+\varphi_B)} dl + z(t).
    \end{align}

    Solving the integral, we have
    \begin{align}\label{eq:S_HRR8}
        y_i(t,\tau) = \gamma_i \mathrm{e}^{\mathrm{j}\omega_0\left(t - \frac{R_O-A_B.\frac{l_2+l_1}{2}\cos(\omega t+\varphi_B)}{c}\right)}\frac{l_2-l_1}{2}\cdot\nonumber\\
        \cdot\operatorname{sinc}\left(\scriptstyle\frac{\omega_0}{2c}A_B(l_2-l_1)\cos(\omega t+\varphi_B)\displaystyle\right) + z(t).
    \end{align}

    Defining for each blade $i$:
    \begin{equation}
        \psi_i(t) = A_B \cos(\omega t + \varphi_B(i)),
    \end{equation}
    \begin{equation}
        \Delta R_i^+(t, \tau) = \frac{l_2+l_1}{2} \psi_i(t),
    \end{equation}
    and
    \begin{equation}
        \Delta R_i^-(t, \tau) = \frac{l_2-l_1}{2} \psi_i(t),
    \end{equation}    
    we can sum up the result in \eqref{eq:S_HRR8} for $N_B$ blades, arriving to our high-range resolution model:
    
    \begin{align}\label{eq:SN_HRR}
        y_i(t,\tau) &= \sum_{i=1}^{N_B} \bigg\{ \gamma_i \mathrm{e}^{\mathrm{j}\omega_0\left(t - \frac{R_O-\Delta R_i^+(t, \tau)}{c}\right)}\nonumber\\
        &\scriptstyle\frac{l_2-l_1}{2}\displaystyle \operatorname{sinc}\left(\scriptstyle\frac{\omega_0}{c}.\Delta R_i^-(t, \tau)\displaystyle\right)\bigg\} + z(t).
    \end{align}

    Differently from the case presented in \autoref{subsection:classic_model}, this model represents the fast-time domain signal before demodulation and matched filtering (channel estimation).


\subsection{OFDM WAVEFORM}\label{subsection:OFDM}

   The model from \eqref{eq:base_radar_signal} assumes a simple single-tone waveform with center frequency $f_0$, where $\omega_0 = 2 \pi f_0$. 
   This formulation can be considered a good approximation for radars with narrow bandwidth. 
   However, for broadband systems, this formulation might not properly reproduce the signal features since the same target point scatterer can produce different Doppler contributions due to the different frequency components of the transmitted signal. 
   In ISAC, it is common to use OFDM waveforms, and for some applications, these waveforms can occupy a considerably broad bandwidth. This case deserves special attention and will be analyzed from this point on.

    The $m^{th}$ transmitted symbol of a system with $N$ OFDM subcarriers can be written as        
    \begin{equation}\label{eq:OFDM_tx_1}
        x(\mu,m) = \sum_{n=0}^{N-1} D(n,m) \mathrm{e}^{\mathrm{j} \omega_n (m + \frac{\mu}{N})T},
    \end{equation}
    where $T$ is the time duration of the OFDM symbol, $f_s = \frac{N}{T}$ is the sampling frequency, and $\mu$ and  $m$ are discrete fast-time and slow-time indexes, respectively, so that the total discrete time $t = mT + \frac{\mu}{N}T$. 
    
    Additionally, ${D(n,m)}$ is the complex amplitude for each subcarrier, also called \textit{complex modulation symbol}, given by the modulation technique, e.g., phase-shift keying (PSK). 
    
    Furthermore, in order to have orthogonality, $\omega_n$ is given by
    \begin{equation}\label{eq:omega_n}
        \omega_n = 2 \pi f_n = 2 \pi (f_0 + \frac{n}{T}),
    \end{equation}
    with $f_n$ denoting the central frequency of each subcarrier.


\subsection{BISTATIC OFDM MATHEMATICAL FORMULATION OF THE SINGLE ROTATING PROPELLER}\label{subsection:Model}

    From \eqref{eq:OFDM_tx_1}, using the same logic as in \cite{Qu2019}, the HRR baseband bistatic returns from a point scatterer $P$, with respect to the $m^{th}$ symbol, can be expressed as
    
    \begin{align}\label{eq:OFDM_ry_P}
        y_P^{(b)}(\mu,m) = &\sum_{n=0}^{N-1}\Bigg\{ \gamma_n D(n,m) \mathrm{e}^{\mathrm{j} \omega_n \left(t-\frac{R_P^{(b)}(t)}{c}\right)}\nonumber\\& 
        \cdot \operatorname{rect} \left(\frac{\tau-\frac{R_P^{(b)}(t)}{c}}{\frac{T}{N}} - \frac{1}{2} \right)\Bigg\} + z(t), 
    \end{align}
    with the discrete delay $\tau = \frac{\mu}{N}T$ and discrete time $t=mT + \frac{\mu}{N}T$.

    In the physical sense, {\eqref{eq:OFDM_ry_P}} represents a received pulse, with duration given by the $\operatorname{rect}$ function, delayed by $\frac{R_P^{(b)}(t)}{c}$ with respect to the original transmitted signal, and composed in frequency by $N$ orthogonal carriers, each one carrying a digitally modulated piece of information, given by $D(n,m)$.

    By substituting the values of $\tau$ and $t$, \eqref{eq:OFDM_ry_P} can be rewritten as    
    \begin{align}\label{eq:OFDM_ry_P_2}
        y_P^{(b)}(\mu,m)& = \sum_{n=0}^{N-1}\Bigg\{ \gamma_n D(n,m) \mathrm{e}^{\mathrm{j} \omega_n \left((m + \frac{\mu}{N})T-\frac{R_P^{(b)}(\mu,m)}{c}\right)}\nonumber\\& 
        \cdot \operatorname{rect} \left(\mu-\frac{R_P^{(b)}(\mu,m)}{c\frac{T}{N}} - \frac{1}{2} \right)\Bigg\} + z_{\mu,m} . 
    \end{align}

    If $P$ is a rotating point scatterer, the bistatic range $R_P^{(b)}(t)$ is given by \eqref{eq:R_t}. If $\tau = \frac{\mu}{N}T$ and $t=mT + \frac{\mu}{N}T$, we can rewrite \eqref{eq:R_t} as
    \begin{equation} \label{eq:R_mu_m}
        R_P^{(b)}(m,\mu) \approx R_O - A_B l \cos\left(\omega T(m + \frac{\mu}{N})+\varphi_B\right),
    \end{equation}
    where $R_O$ and $A_B$ should be calculated for each discrete instant, according to the target movement model.

    Once again, considering a blade $i$ as an unidimensional line composed of infinite point scatterers, the entirety of the returns of one blade can be calculated by integrating the returns of all point scatters using the following equation:

    \begin{align}
        \resizebox{\linewidth}{!}{$
        \begin{aligned}
            y_i^{(b)}(\mu,&m) = \int_{0}^{L_B} \sum_{n=0}^{N-1} \Bigg\{ \gamma_n D(n,m) \mathrm{e}^{\mathrm{j} \omega_n \left[(m + \frac{\mu}{N})T-\frac{R_P^{(b)}(\mu,m)}{c}\right]}
            \\& \cdot \operatorname{rect} \left(\mu-\frac{R_P^{(b)}(\mu,m)}{c\frac{T}{N}} - \frac{1}{2} \right) \Bigg\} \, dl + z_{\mu,m}.
        \end{aligned}
        $}
        \label{eq:OFDM_rx_l}
    \end{align}
    
    If $l_1$ and $l_2$ correspond to the limits of the blade length comprised by the pulse length in the delay $\tau = \frac{\mu}{N}T$, then

    \begin{align}\label{eq:OFDM_rx_l_1}
    y_i^{(b)}(\mu,m) = \sum_{n=0}^{N-1}\Bigg\{ \gamma_n D(n,m) \mathrm{e}^{\mathrm{j} \omega_n T (m + \frac{\mu}{N})} \nonumber \\ \int_{l1}^{l2}{\mathrm{e}^{-\mathrm{j} \frac{\omega_n}{c} \left[R_P^{(b)}(m,\mu)\right]}}{dl}\Bigg\} + z_{\mu,m}.
    \end{align}

    By applying \eqref{eq:omega_n} and \eqref{eq:R_mu_m}, we have
    \begin{align}\label{eq:OFDM_rx_l_2}
    &y_i^{(b)}(\mu,m) = \mathrm{e}^{\mathrm{j} \omega_0 T\left(m + \frac{\mu}{N}\right)} \sum_{n=0}^{N-1}\Bigg\{ \gamma_n D(n,m) \mathrm{e}^{\mathrm{j} 2\pi n \left(m + \frac{\mu}{N}\right)} \nonumber \\&\int_{l1}^{l2}{\mathrm{e}^{-\mathrm{j} \frac{\omega_n}{c} \left[R_O - A_B.l\cos\left(\omega T(m + \frac{\mu}{N})+\varphi_B\right)\right]}}{dl}\Bigg\} + z_{\mu,m}.
    \end{align}

    The returns of all $N_B$ blades, with $\varphi_B$ given, similarly to \eqref{eq:phi_B_i}, by $\varphi_B(i) = \varphi_B + 2\pi\frac{i-1}{N_B}$, can be summed with
    
    \begin{equation}\label{eq:OFDM_rx_l_3}
        y^{(b)}(\mu,m) = \sum_{i=1}^{N_B}\Bigg\{ y_i^{(b)}(\mu,m) \Bigg\}.
    \end{equation}

    Following the same steps as in \autoref{subsection:high_resolution_bistatic}, we solve the integral, calculate its limits, and eliminate the carrier frequency component. 
    As a result, the proposed model in the time domain is concluded by describing the baseband signal as
    \begin{align}\label{eq:final_propeller_model}
        y^{(b)}(\mu, m) = \sum_{n=0}^{N-1} \Bigg\{ D(n,m) \mathrm{e}^{\mathrm{j} 2 \pi \frac{n \mu}{N}}\nonumber
        \\\cdot\sum_{i=1}^{N_B} \biggl[ \gamma_{ni}\mathrm{e}^{\mathrm{j}\frac{\omega_n}{c}\left(-R_O+\Delta R_i^+(\mu)\right)}\nonumber
        \\\cdot\frac{l_2-l_1}{2} \operatorname{sinc}\left(\frac{\omega_n}{c} \Delta R_i^-(\mu)\right)\biggr] \Bigg\} + z_{\mu,m},
    \end{align}
    where 
    \begin{equation}
        \Delta R_i^+(\mu, m) = \frac{l_2+l_1}{2} \psi_i(\mu, m),
    \end{equation}
    
    \begin{equation}
        \Delta R_i^-(\mu, m) = \frac{l_2-l_1}{2} \psi_i(\mu, m),
    \end{equation}
    and    
    \begin{equation}
        \psi_i(\mu, m) = A_B \cos\left(\omega T (m + \frac{\mu}{N}) + \varphi_B(i)\right).
    \end{equation}
    With $l_1$ and $l_2$ being the medians, respectively, of the following sets:
        
    \begin{equation}\label{eq:l1_OFDM_2}
        \bigg\{0, \frac{R_O - \frac{\mu-1}{N} c T}{\psi_i(\mu, m)}, \frac{R_O - \frac{\mu}{N}c T}{\psi_i(\mu, m)}\bigg\},
    \end{equation}
    
    \begin{equation}\label{eq:l2_OFDM_2}
        \bigg\{L_B, \frac{R_O - \frac{\mu-1}{N} c T}{\psi_i(\mu, m)}, \frac{R_O - \frac{\mu}{N} c T}{\psi_i(\mu, m)}\bigg\}.
    \end{equation}

    Therefore, \eqref{eq:final_propeller_model} represents the proposed OFDM bistatic micro-Doppler model for a single rotating propeller.
    
    An important contribution of this proposed model is that it is two-dimensional (delay and slow-time), which gives this model the ability to generate high-range resolution signatures, which differs from the classic thin wire model for the micro-Doppler of rotating blades presented in \cite{Schneider1988} and widely used in the literature, that assumes the target response as completely contained in one range resolution cell.
    
\subsection{VALIDATION OF THE SINGLE ROTATING PROPELLER MODEL}
    
    The model described in \eqref{eq:final_propeller_model} was validated against the measurements detailed in Appendix \ref{subsection:BiRa_measurement}, carried out using the BIRA Measurement System, described in Appendix \ref{subsection:BiRa}.

    In the current work, Doppler frequency analysis will be used to validate the model for low-resolution cases, while range-Doppler analysis will be used for validation in the HRR scenario since the range dimension only becomes important in this case.

\subsubsection{LOW RANGE RESOLUTION CASE}

\begin{figure}[t]
    \centering
    \includegraphics[width=0.45\textwidth, height=2in]{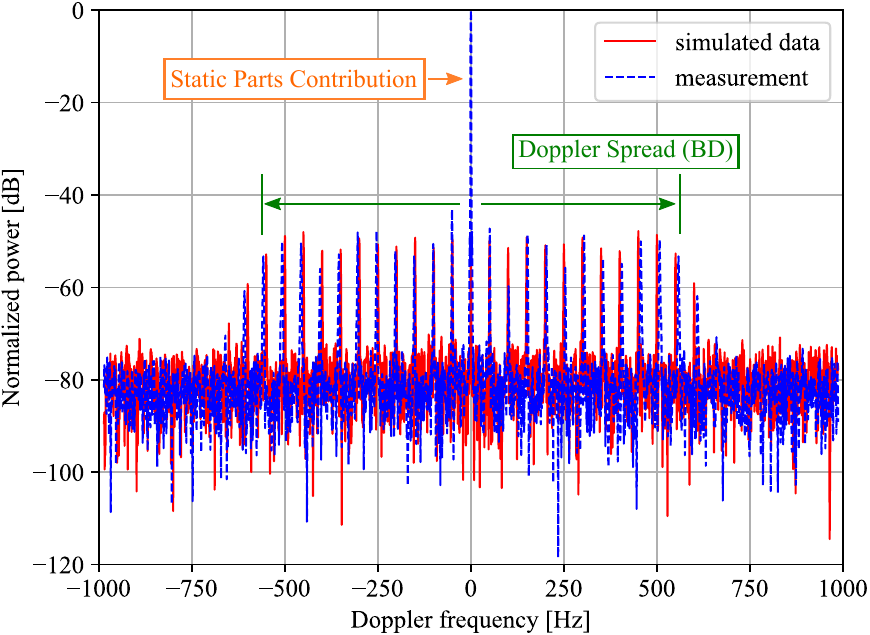}
    \caption{Comparison between simulation and measurement micro-Doppler signatures of a rotating propeller (16384 symbols, subsampling = 8, $\beta = \ang{30}$).}
    \label{fig:simul_measurement}
\end{figure}

    \autoref{fig:simul_measurement} presents a comparison between two frequency domain micro-Doppler signatures of a single rotating propeller: one simulated using the model developed in Sections \ref{section:Model} and \ref{section:Drone_Model}, and one obtained from the measurements. 
    Both use the parameters described in Table \ref{tab_MicroDoppler_measurement_setup} Setup 1, in the Appendix \ref{subsection:BiRa_measurement}, and a geometric configuration with bistatic angles of \ang{60}.   
    The OFDM coefficients $D(n,m)$ used in this simulation were obtained from the transmitted signal employed in the measurement.
    
    As explained in \cite{costa2024bistatic}, the frequency domain signature of rotating propellers is a periodic sequence of spikes separated by $\Delta f = N_B \omega/2\pi$ and distributed over a Doppler spread of $B_D = 4 \omega L_B \cos{(\beta/2)}\sin(\psi)/\lambda_0$.    
    As shown in \autoref{fig:simul_measurement}, the simulated signature matches well the response obtained from the measured data, presenting the same Doppler spread and the same distance between the peaks, which are the most important features for target classification.
    The main difference that can be seen between the two signatures is a strong peak in the center, referring to the contributions of the static parts of the drone, which are not modeled by \eqref{eq:final_propeller_model}.
    
\begin{figure}[t]
        \centering
        \includegraphics[width=0.45\textwidth, height=1.75in]{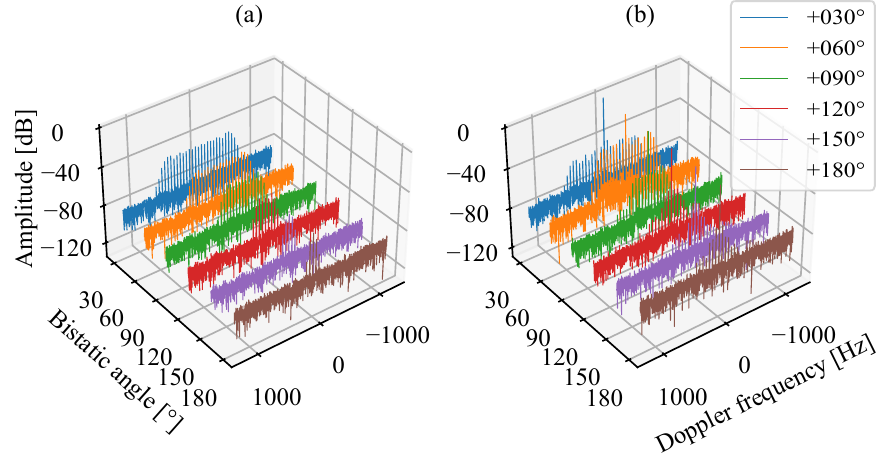}        
        \caption{Comparison of frequency-domain micro-Doppler signatures of a rotating propeller from: (a) proposed model;  (b) measurement data, for different bistatic angles (16384 symbols, subsampling = 8). The legend indicates the bistatic angle.}
        \label{fig:bistatic_angles_simul}
    \end{figure}

    Beyond that, the comparison between \autoref{fig:bistatic_angles_simul}a and \autoref{fig:bistatic_angles_simul}b shows that the proposed model is also capable of emulating the dependence on the bistatic angle seen in the measurement data. 
    
    Finally, the Pearson correlation coefficient was used to check the similarity between the frequency domain micro-Doppler signatures produced by the proposed model and by measurement.
    This produces a cross-correlation coefficient of 0.98 across the bistatic angles from \ang{30} to \ang{180}.
    
    
\subsubsection{HIGH RANGE RESOLUTION CASE}
    
\begin{figure}[t]
    \centering
    \includegraphics[width=0.45\textwidth, height=2in]{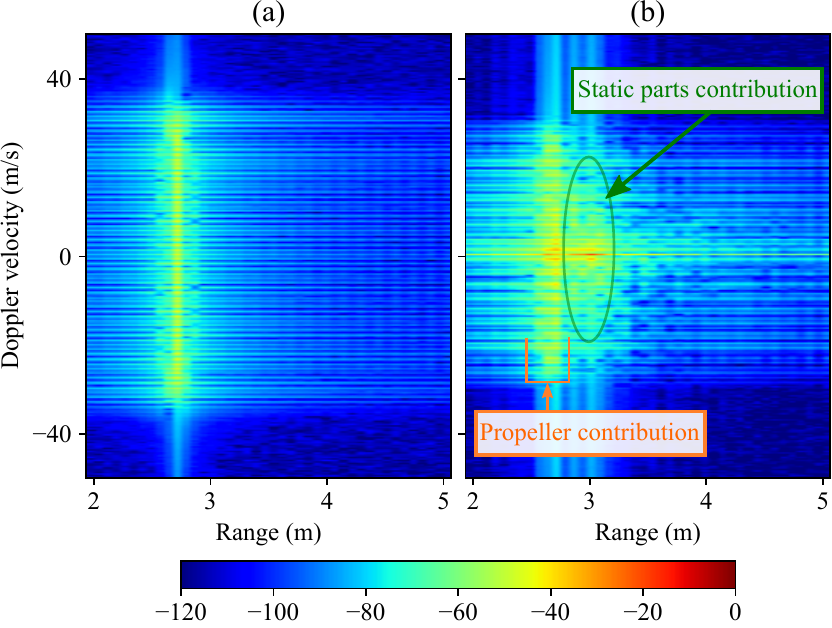}
    \caption{High Range Resolution comparison between (a) simulation and (b) measurement micro-Doppler signatures of a rotating propeller ($\beta = \ang{10}$).}
    \label{fig:HRR_propeller_comparison}
\end{figure}

    \autoref{fig:HRR_propeller_comparison} shows a comparison between the range-Doppler signatures of a rotating propeller from data simulated using the proposed model and obtained from measurements of a drone with just one active propeller. 
    Both the simulation and the real measurement use the parameters from Table \ref{tab_MicroDoppler_measurement_setup} Setup 2, which represents a high-range resolution scenario.
    It can be seen that the main properties of the propeller micro-Doppler signatures for the sake of classification, namely the format and Doppler spread of the propeller, are similar on both sides.
    As well as in \autoref{fig:simul_measurement}, the main difference is that the measurement presents the contribution of the drone static parts.
    
\section{EXTENSION TO THE MICRO-DOPPLER MODEL OF A MULTI-PROPELLER DRONE}\label{section:Drone_Model}
In \autoref{section:Model}, a mathematical model for a single rotating propeller was derived. 
Now, in order to fully simulate a drone, it is necessary to include two main contributions.

The first contribution is the combined signature of all of its rotating propellers.
This contribution is mainly responsible for the micro-Doppler signature of the target.
The second is the contribution from the static parts of the drone.

From these two contributions, we can derive the total returns from the drone by the following:
    \begin{equation}\label{eq:drone_model}
        y^{(b)}(\mu, m) = \sum_{p=1}^{N_p}{\big\{{y_p}^{(b)}(\mu, m)\big\}} + {y_s}^{(b)}(\mu, m),
    \end{equation}
where $y_p$ stands for the total returns due to the propellers, $N_p$ is the number of propellers in the drone, and $y_s$ represents the returns from static parts.

The combination of multiple propellers and the contribution of static parts will be addressed in the next subsections.

\subsection{COMBINING PROPELLERS} \label{subsection:Multiple_Propellers}

    
    \begin{figure}[b]
        \centering
        \includegraphics[width=0.45\textwidth]{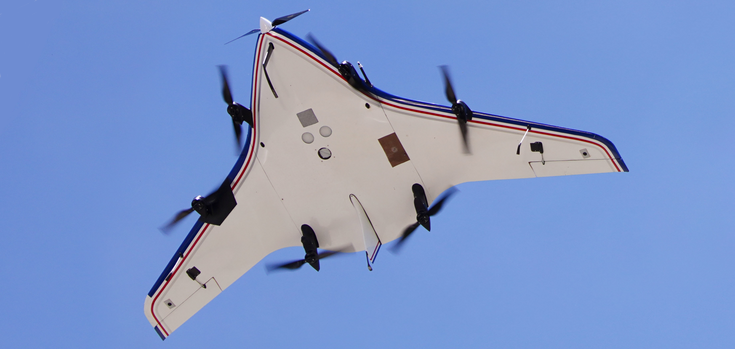}
        \caption{VTOL with 6 horizontal propellers and 1 vertical propeller.}
        \label{fig:VTOL}
    \end{figure}

    \subsubsection{Extension of the Single Propeller Model}
    
    The geometry of the drone will be used to properly represent the propeller locations.
    The position vectors of the transmitter $\boldsymbol{\vec{T}}$, the receiver $\boldsymbol{\vec{R}}$, and the drone center $\boldsymbol{\vec{O_d}}$ are the references for all propellers.
    Then, in a drone with $N_P$ propellers, for each propeller $p \in \{1,...,N_P\}$, the propeller rotation center $\boldsymbol{\vec{O}}$ becomes $\boldsymbol{\vec{O}}_p$, the angular velocity vector $\boldsymbol{\vec{\omega}}$ becomes $\boldsymbol{\vec{\omega}}_p$, as well as the initial angular position ${\varphi_0}$ from \eqref{eq:phi_B_i} becomes ${\varphi_0}_p$, to calculate the parameters $R_O$, $\beta_T$, and $\beta_R$, by doing:
    
    \begin{equation}
        \omega_p = \|\boldsymbol{\vec{\omega}}_p\|,
    \end{equation}
    
    \begin{equation}
        {R_O}_p = \|\boldsymbol{\vec{OT}}_p\|,
    \end{equation}
    
    \begin{equation}
        {\beta_T}_p = \operatorname{arccos}\left(\frac{\boldsymbol{\vec{\omega}}_p\cdot\boldsymbol{\vec{OT}}_p}{\|\boldsymbol{\vec{\omega}}_p\cdot\boldsymbol{\vec{OT}}_p\|}\right),
    \end{equation}
    and
    \begin{equation}
        {\beta_R}_p = \operatorname{arccos}\left(\frac{\boldsymbol{\vec{\omega}}_p\cdot\boldsymbol{\vec{OR}}_p}{\|\boldsymbol{\vec{\omega}}_p\cdot\boldsymbol{\vec{OR}}_p\|}\right),
    \end{equation}
    where $\boldsymbol{\vec{OT}}_p = \boldsymbol{\vec{O}}_p-\boldsymbol{\vec{T}}$ and $\boldsymbol{\vec{OR}}_p = \boldsymbol{\vec{O}}_p-\boldsymbol{\vec{R}}$.

    Now, these extended parameters are applied to \eqref{eq:final_propeller_model} to generate the returns $y_p$ of each propeller, and applied to \eqref{eq:drone_model} to generate the multiple propellers model.
    
    
    This formulation is very flexible, making it possible to place propellers independently in any position and orientation.
    Furthermore, propellers usually have random initial azimuth angles (${\varphi_0}_p = \mathcal{U}(0,2\pi)$) and different rotation speeds $\omega_p$, which can also be easily implemented with the proposed formulation.
    Therefore, we can not only create the most common drone models but also simulate more complex systems, such as vertical take-off and landing (VTOL) drones, as shown in \autoref{fig:VTOL}, that have a propeller spinning in a direction orthogonal to all other propellers.
    
    
\begin{figure}[t]
    \centering
    \includegraphics[width=0.45\textwidth, height=2in]{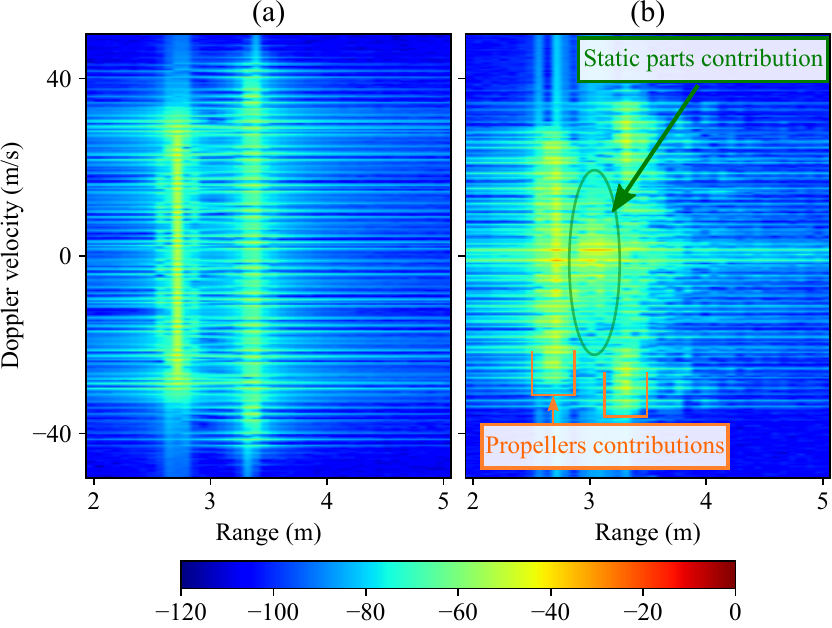}
    \caption{HRR comparison between (a) simulation and (b) measurement micro-Doppler signatures of two rotating propellers with different rotation speeds ($\beta = \ang{10}$).}
    \label{fig:hrr_2propellers_comparison}
\end{figure}

    \subsubsection{Validation of the Multiple Propellers Model}

    \autoref{fig:hrr_2propellers_comparison} depicts a comparison between the HRR range-Doppler signature of two rotating propellers, generated using the method proposed in the current subsection, and measurement data of a drone with two active propellers.
    All configurations, including the geometry, are the same on both sides.
    Two rotation speeds (\qty{1500}{rpm} and \qty{2000}{rpm}) were applied to the propellers in each case, which explains the difference in the Doppler spread inside each figure.
    One more time, it can be seen that the propeller contributions have similar shapes, and Doppler spreads according to the rotation speed of each propeller. 
    Additionally, corresponding propellers appear in the same ranges.

    Evidently, the contribution of the drone body is still missing in the simulation, which takes us to the next step.

\subsection{ADDING BODY SIGNATURE} \label{subsection:Static}

Most of the formulations that adopt the thin wire model for propellers use a single-point scatterer, usually located in the center of the drone, to stand for the static parts returns. 
This method gives good results for narrowband cases, where all returns from the target sum up to a single range resolution bin. 
However, for broadband cases, it is necessary to reproduce the contributions of the drone body frame returns to different range bins.

In this work, we present two approaches to generate reflections of these distributed static parts returns. 
    
    \begin{figure*}[t]
        \centering
        \includegraphics[width=0.95\textwidth]{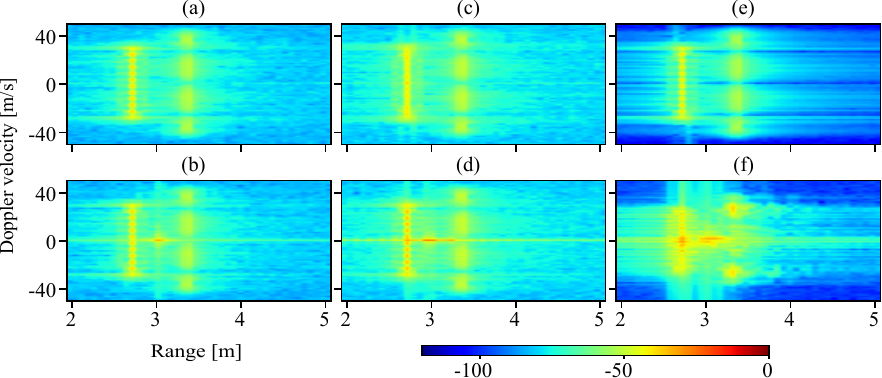}
        \caption{Comparison of different body models using HRR range-Doppler signatures of a two propellers drone ($\boldsymbol{\beta} \boldsymbol{=} \mathbf{10^{\circ}}$).
                \\Gaussian body model: (a) without vibration, (b) with vibration;
                \\Measurement-based body model: (c) without vibration, (d) with vibration;
                \\Other cases for reference: (e) simulation with no body model; and (f) measurement (includes drone body and vibration).}
        \label{fig:comparison_HRR}
    \end{figure*}

\subsubsection{Gaussian Body Model}
The first is a simple extension of the single-point scatterer method. 
A Gaussian function, centered on the mass center of the target, was used to spread the drone body frame energy over the range bins.
The Gaussian body model assumes a smoothly distributed RCS centered on the drone's mass center.
This is clearly a simplification, as real drone bodies exhibit intricate reflectivity patterns that depend on the shape, material and aspect angle.
This simplification allows a simple representation of the drone body as an extended target in the wideband range-Doppler profile.

A Gaussian centered on the drone's mass center $R^{(b)}$ and standard deviation equal to the size of the target can be written as
\begin{gather}
    \operatorname{g}(t, \tau) = \frac{1}{d'_\text{max}\sqrt{2\pi}}e^{-\frac{1}{2}\frac{(c\tau-R^{(b)}(t))^2}{(d'_\text{max})^2}},\label{eq:gaussian_function}
\end{gather}
where $d'_\text{max} = d_\text{max} \cos{\frac{\beta}{2}}$, $d_\text{max}$ is the maximum dimension of the target in the direction of the bistatic bisector, and recall that $\tau = \frac{\mu}{N}T$ and $t=mT+\frac{\mu}{N}T$ are, respectively, the discrete delay and discrete time.

A necessary care is necessary in forward scattering case, since in this situation $\cos(\beta/2) = 0$, what causes an indetermination in the Gaussian function.
To solve this, we can just remember that the minimum relevant dimension is the bistatic range resolution.
Therefore, we can use  $d'_\text{max} = \max\left(d_\text{max} \cos{\frac{\beta}{2},\frac{\mu}{N} T c}\right)$.

Applying $\gamma_n = \gamma'_n \operatorname{g}(t, \tau) d'_\text{max}\sqrt{2\pi}$ to \eqref{eq:OFDM_ry_P_2}, scaling the $\operatorname{rect}$ function by the size of the target, and substituting the values of $t$ and $\tau$, we write the Gaussian Body Model as

\begin{gather}
    {y_s}^{(b)}(\mu, m) = e^{-\frac{1}{2} \frac{(\frac{\mu T c}{N}-R^{(b)}(\mu,m))^2}{(d'_\text{max})^2}}\nonumber\\ \cdot\operatorname{rect}\Bigg({\scriptstyle\frac{c\frac{T}{N}}{d_\text{max} \cos{\frac{\beta}{2}}}\left(\mu-\frac{R^{(b)}(\mu,m)}{c\frac{T}{N}} - \frac{1}{2}\right)\displaystyle}\Bigg)\nonumber\\
    \cdot \sum_{n=0}^{N-1}\Bigg[ \gamma_n D(n,m) \mathrm{e}^{\mathrm{j} \omega_n \left(mT + \frac{\mu T}{N}-\frac{R^{(b)}(\mu,m)}{c}\right)}\Bigg] + z_{\mu,m},\label{eq:body_normal_model}
\end{gather}
where $\gamma'_n$ defines the body reflectivity with respect to the propellers reflectivity.
The appropriate values of $\gamma'_n$ depend strongly on the material of the drone and, for simplicity, in the current work, they are considered equal for all values of $n$, and this single value was manually adjusted, similarly to what is usually done in the single-point scatter method.
    
\subsubsection{Measurement-based Body Model}
The second strategy carried out was to use static reflectivity values obtained from measurements. 
The idea here is to perform high-resolution bistatic reflectivity measurements of the drone in multiple aspect angles, in order to create a library with range reflectivity profiles $S_{21}(f)$ of the drone static parts.

According to the desired geometry, we select the appropriate reflectivity profile from the library, and sample $S_{21}(f)$ by $f_s = \frac{N}{T}$, in order to get the discrete reflectivity profile $S'_{21}(n) = S_{21}(f_n)$, where $f_n$ is the same as in \eqref{eq:omega_n}.

Then, applying $\gamma_n = \gamma'_n S'_{21}(n)$ to \eqref{eq:OFDM_ry_P_2}, the Measurement-based Body Model can written as
\begin{gather}
{y_s}^{(b)}(\mu, m) = \sum_{n=0}^{N-1}\Bigg[ \gamma_n S'_{21}(n) D(n,m)\nonumber \\ \cdot \mathrm{e}^{\mathrm{j} \omega_n \left(mT + \frac{\mu T}{N}-\frac{R^{(b)}(\mu,m)}{c}\right)}\nonumber\\ 
\cdot \operatorname{rect}\Bigg({\scriptstyle\frac{c\frac{T}{N}}{d_\text{max} \cos{\frac{\beta}{2}}}\left(\mu-\frac{R^{(b)}(\mu,m)}{c\frac{T}{N}} - \frac{1}{2}\right)\displaystyle}\Bigg)\Bigg] + z_{\mu,m},\label{eq:body_measurement_model}
\end{gather}


Details on how these measurements $S_{21}(f)$ were performed are provided in Appendix \ref{section:Static_measurements}.

When measurements are not available, the static reflectivity of the drone body can also be generated using full-wave electromagnetic (EM) simulation.

\subsubsection{Vibration Micro-Doppler Contribution}

Additionally, as can be seen in \autoref{fig:comparison_HRR}f, the contribution of the static parts also presents some Doppler spread. 
That happens due to the vibration of the target, making the range-Doppler signature in this case clearly distinct from vibrationless models, as in \autoref{fig:comparison_HRR}a and \autoref{fig:comparison_HRR}c.

Micro-Doppler due to vibration can be emulated by adding a time-dependent change in phase due to fast short-distance displacements.
\cite{chen2014radar} has modeled these displacements as sinusoids. 
In this work, a displacement between two consecutive time steps $t-1$, and $t$ was modeled as an uniformly distributed random variable, with maximum displacement due to vibration given by $D_0$, as shown in \eqref{eq:vibration_mD} and \eqref{eq:vibration_mD_2}:
\begin{equation}\label{eq:vibration_mD}
    R^{(b)}_v(t) = R^{(b)}(t) + D_v(t),
\end{equation}
where
\begin{equation}\label{eq:vibration_mD_2}
    D_v(t) = D_v(t-1) + D_0\cdot\mathcal{U}(-1,1).
\end{equation}

\section{VALIDATION OF THE PROPOSED MULTI-PROPELLER DRONE MODEL}  \label{section:validation} 
    \autoref{fig:comparison_HRR} presents a comparison between the four different methods for generating the contribution of the drone body.
    \autoref{fig:comparison_HRR}a and \autoref{fig:comparison_HRR}c show the drone simulation using the Gaussian body model and measurement-based body model, respectively, without considering vibration.
    The figures show that in both cases, the contribution of the body is reduced to zero Doppler.
    In that case, it is hard to see this contribution in the HRR analysis and, therefore, there is not much improvement when compared to the bodiless model, as in \autoref{fig:hrr_2propellers_comparison}a and \autoref{fig:comparison_HRR}e.

    In \autoref{fig:comparison_HRR}b and \autoref{fig:comparison_HRR}d, the Gaussian and measurement-based models with vibration are illustrated, respectively.
    In these cases, it can be seen that the model is able to fairly represent the contributions of the static parts since they present a Doppler spread around the zero Doppler line and with more energy around the body position in a way similar to what can be seen in \autoref{fig:comparison_HRR}f, for the measurement case.    

    \pgfplotstableread[col sep=comma]{costa8.csv}\datatable
    \begin{figure}[t]
        \centering
        \begin{tikzpicture}
\tikzstyle{every node}=[font=\small]
\begin{groupplot}[
  group style={
    group size=2 by 2,
    horizontal sep=0.5cm,
    vertical sep=0.5cm,
  },
  width=0.275\textwidth,
  height=0.2\textwidth,
  xlabel={Bistatic Angle [°]},
  xtick=data,
  grid=both,
  yticklabel style={
    /pgf/number format/fixed,
    /pgf/number format/precision=2
  },
  legend style={
    at={(1,-0.7)},
    anchor=south east,
    font=\small,
    draw=none,
    cells={align=left},
  },
  legend columns=3
]

\nextgroupplot[
  title={Comparison of \\ [-0.8ex] complex values},ylabel={MSE}, ymin=0, ymax=0.6,
  title style = {align = center, font=\small, yshift=-1ex},
  xticklabels=\empty,
  xlabel={}]
\addplot+[black, mark=x, thick, mark options={fill=none, draw=black, scale=2}] table[x={Bistatic Angle [°]}, y={MSE2}] {\datatable};
\addplot+[blue, mark=+, thick, mark options={fill=none, draw=blue, scale=2}] table[x={Bistatic Angle [°]}, y={MSE3}] {\datatable};
\addplot+[red, mark=o, thick, mark options={fill=red}] table[x={Bistatic Angle [°]}, y={MSE}] {\datatable};

\nextgroupplot[
  title={Comparison of \\ [-0.8ex] magnitude}, ymin=0, ymax=0.6,
  title style = {align = center, font=\small, yshift=-1ex},
  xticklabels=\empty,
  xlabel={},
  yticklabels=\empty,
  yticklabel pos=right]
\addplot+[black, mark=x, thick, mark options={fill=none, draw=black, scale=2}] table[x={Bistatic Angle [°]}, y={MSE2_abs}] {\datatable};
\addplot+[blue, mark=+, thick, mark options={fill=none, draw=blue, scale=2}] table[x={Bistatic Angle [°]}, y={MSE3_abs}] {\datatable};
\addplot+[red, mark=o, thick, mark options={fill=red}] table[x={Bistatic Angle [°]}, y={MSE_abs}] {\datatable};

\nextgroupplot[
  ylabel={Correlation coefficient}, ymin=0, ymax=1]
\addplot+[black, mark=x, thick, mark options={fill=none, draw=black, scale=2}] table[x={Bistatic Angle [°]}, y={ Correlation Coefficient2}] {\datatable};
\addplot+[blue, mark=+, thick, mark options={fill=none, draw=blue, scale=2}] table[x={Bistatic Angle [°]}, y={ Correlation Coefficient3}] {\datatable};
\addplot+[red, mark=o, thick, mark options={fill=red}] table[x={Bistatic Angle [°]}, y={Correlation Coefficient}] {\datatable};

\nextgroupplot[
  ymin=0, ymax=1,
  yticklabels=\empty,
  yticklabel pos=right]
\addplot+[black, mark=x, thick, mark options={fill=none, draw=black, scale=2}] table[x={Bistatic Angle [°]}, y={ Correlation Coefficient2 (abs)}] {\datatable};
\addplot+[blue, mark=+, thick, mark options={fill=none, draw=blue, scale=2}] table[x={Bistatic Angle [°]}, y={ Correlation Coefficient3 (abs)}] {\datatable};
\addplot+[red, mark=o, thick, mark options={fill=red}] table[x={Bistatic Angle [°]}, y={Correlation Coefficient (abs)}] {\datatable};

\addlegendentry{Discrete points}
\addlegendentry{Single-point}
\addlegendentry{Proposed model}

\end{groupplot}
\end{tikzpicture}
        \caption{Mean squared error (MSE) and Pearson correlation coefficient between frequency domain micro-Doppler signatures of models and measurement (1 propeller with Gaussian body, 16384 symbols, subsampling = 8).}
        \label{fig:mse_1propeller_gaussian}
    \end{figure}
    
    The measurement-based model has the advantage of reproducing the range-dependent distribution of the target better, resulting in a more accurate model. 
    However, it has two obvious disadvantages. 
    The first one is that it requires measurements, or at least full-wave EM simulation, for each target to be modeled. 
    The other is that it is unfeasible to perform static reflectivity measurements for all possible geometric constellations, i.e., all possible combinations of bistatic angles. 

    In contrast, the Gaussian body model has the advantages of not needing measurements and the possibility of use with any geometric constellation while having the disadvantage of being less accurate than the measurement-based model.
    Therefore, the Gaussian or measurement-based model should be chosen according to the possibilities and requirements.

    \autoref{fig:mse_1propeller_gaussian} shows the mean squared error (MSE) and the Pearson correlation coefficient between models and the measurement for different bistatic angles. 
    Beside the proposed model, two other models were used for comparison, namely the single-point scatterer model and the discrete points model.
    In the proposed model, the Gaussian body model was used in this analysis, since it is easier to be reproduced by the reader. 
    Just one propeller was used, in order to balance the weight between propeller and body model in the analysis.
    As input to calculate the MSE and the correlation coefficient, we used, at the left-hand side and right-hand side of the figure, respectively, the complex and the magnitude values of the frequency-domain micro-Doppler signature, normalized between 0 and 1, what gives a scale between 0 and 1 for both metrics.
    
    The Pearson correlation coefficient $\big( r = \frac{\sum_{i=1}^n (x_i - \bar{x})(y_i - \bar{y})}{\sqrt{\sum_{i=1}^n (x_i - \bar{x})^2} \sqrt{\sum_{i=1}^n (y_i - \bar{y})^2}}\big)$ is a scale invariant metric for comparisons, while the MSE $(\mathrm{mse} = \frac{1}{n} \sum_{i=1}^{n} (y_i - x_i)^2)$ shows an absolute comparison. Here, $y_i$ comes from the model, $x_i$ comes from the measurement, and $\bar{y}$ and $\bar{x}$ are the mean of all values in $y_i$ and $x_i$, respectively.
    
    The results show that for frequency-domain ({\autoref{fig:mse_1propeller_gaussian}}) all models achieve low MSE values (under 0.01), and high Pearson correlation coefficients (above 0.99), with no significative difference between then.
    However, for range-Doppler representation ({\autoref{fig:rdmap_mse_1propeller_gaussian}}), while the other models cannot achieve good results, simply because they were not created to support range-Doppler analysis, the proposed model maintains very good performance.

    For this comparison, the single-point model used in \mbox{\cite{chen2000time}}, \mbox{\cite{chen2006micro}}, \mbox{\cite{White2022}}, and \mbox{\cite{Wei2024}}, was adapted to bistatic geometry, since the originals are all monostatic.
    Likewise, the discrete points model, that has a bistatic version presented by \mbox{\cite{Plotnitskaya2021}}, was also adapted, since its original description supports only the case where the angular velocity vector $\boldsymbol{\vec{\omega}}_p$ is orthogonal to the baseline $\boldsymbol{TxRx}$.

    \pgfplotstableread[col sep=comma]{costa9.csv}\datatable
    \begin{figure}[t]
        \centering
        \begin{tikzpicture}
\tikzstyle{every node}=[font=\small]
\begin{groupplot}[
  group style={
    group size=2 by 2,
    horizontal sep=0.5cm,
    vertical sep=0.5cm,
  },
  width=0.275\textwidth,
  height=0.2\textwidth,
  xlabel={Bistatic Angle [°]},
  xtick=data,
  grid=both,
  yticklabel style={
    /pgf/number format/fixed,
    /pgf/number format/precision=2
  },
  legend style={
    at={(1,-0.7)},
    anchor=south east,
    font=\small,
    draw=none,
    cells={align=left},
  },
  legend columns=3
]

\nextgroupplot[
  title={Comparison of \\ [-0.8ex] complex values},ylabel={MSE}, ymin=0, ymax=0.6,
  title style = {align = center, font=\small, yshift=-1ex},
  xticklabels=\empty,
  xlabel={}]
\addplot+[black, mark=x, thick, mark options={fill=none, draw=black, scale=2}] table[x={Bistatic Angle [°]}, y={MSE2}] {\datatable};
\addplot+[blue, mark=+, thick, mark options={fill=none, draw=blue, scale=2}] table[x={Bistatic Angle [°]}, y={MSE3}] {\datatable};
\addplot+[red, mark=o, thick, mark options={fill=red}] table[x={Bistatic Angle [°]}, y={MSE}] {\datatable};

\nextgroupplot[
  title={Comparison of \\ [-0.8ex] magnitude}, ymin=0, ymax=0.6,
  title style = {align = center, font=\small, yshift=-1ex},
  xticklabels=\empty,
  xlabel={},
  yticklabels=\empty,
  yticklabel pos=right]
\addplot+[black, mark=x, thick, mark options={fill=none, draw=black, scale=2}] table[x={Bistatic Angle [°]}, y={MSE2_abs}] {\datatable};
\addplot+[blue, mark=+, thick, mark options={fill=none, draw=blue, scale=2}] table[x={Bistatic Angle [°]}, y={MSE3_abs}] {\datatable};
\addplot+[red, mark=o, thick, mark options={fill=red}] table[x={Bistatic Angle [°]}, y={MSE_abs}] {\datatable};

\nextgroupplot[
  ylabel={Correlation coefficient}, ymin=0, ymax=1]
\addplot+[black, mark=x, thick, mark options={fill=none, draw=black, scale=2}] table[x={Bistatic Angle [°]}, y={ Correlation Coefficient2}] {\datatable};
\addplot+[blue, mark=+, thick, mark options={fill=none, draw=blue, scale=2}] table[x={Bistatic Angle [°]}, y={ Correlation Coefficient3}] {\datatable};
\addplot+[red, mark=o, thick, mark options={fill=red}] table[x={Bistatic Angle [°]}, y={Correlation Coefficient}] {\datatable};

\nextgroupplot[
  ymin=0, ymax=1,
  yticklabels=\empty,
  yticklabel pos=right]
\addplot+[black, mark=x, thick, mark options={fill=none, draw=black, scale=2}] table[x={Bistatic Angle [°]}, y={ Correlation Coefficient2 (abs)}] {\datatable};
\addplot+[blue, mark=+, thick, mark options={fill=none, draw=blue, scale=2}] table[x={Bistatic Angle [°]}, y={ Correlation Coefficient3 (abs)}] {\datatable};
\addplot+[red, mark=o, thick, mark options={fill=red}] table[x={Bistatic Angle [°]}, y={Correlation Coefficient (abs)}] {\datatable};

\addlegendentry{Discrete points}
\addlegendentry{Single-point}
\addlegendentry{Proposed model}

\end{groupplot}
\end{tikzpicture}
        \caption{Mean squared error (MSE) and Pearson correlation coefficient between HRR (range-Doppler) micro-Doppler signatures of models and measurement (1 propeller with Gaussian body, 16384 symbols, subsampling = 8).}
        \label{fig:rdmap_mse_1propeller_gaussian}
    \end{figure}

\section{ADVANTAGES AND DISADVANTAGES OF THE PROPOSED MODEL}\label{section:AdvantagesDisadvantages}

    As mentioned in the Introduction, various other methods have been employed to model micro-Doppler.
    Some authors have used CAD-based EM simulation methods, such as the Finite Element Method (FEM), Methods of Moments (MoM), and Physics Optics (PO).
    Compared to these methods, the proposed model tends to be less accurate since it does not take into account some electromagnetic phenomena, such as diffraction and multipath.
    On the other hand, the most important advantages of the proposed method are simplicity and low computational cost.
    Its simplicity makes it suitable for applications where a large amount of simulated data with various parameters is required.
    
    Some other authors have used modeling based on single-point scatterer, finite discrete point scatterers, or the classic monostatic narrowband thin wire model. 
    Those methods can potentially have a lower computational cost than the proposed one. 
    However, the proposed model also has a very low computational cost.
    Moreover, the proposed method tends to be more accurate and has the ability to represent the HRR case while taking into account the bistatic geometry and the OFDM waveform.

    This model also has another considerable advantage when compared to CAD-based EM simulation methods. 
    It can easily reproduce the drone fast movements and rotations, with 3D velocity and 3D angular changes, which appear in practical scenarios, while vibration can be included with {\eqref{eq:vibration_mD}}.

    In the proposed model, Doppler information cannot be directly inserted or obtained. 
    The model only calculates the returns received at each instant, considering the position of the point scatters at that moment.
    Velocities and Doppler information are mere consequences of those positions.
    
    As we can see in \eqref{eq:final_propeller_model} and \eqref{eq:body_normal_model}, each slow-time sample generated depends only on the rotation center ranges $R_O(t)$, on the range of the center of the drone body $R^{(b)}(t)$, and on the angular velocity vector for each propeller $\boldsymbol{\vec{\omega}}_p$, which carries 3D angular information of the target.
    These ranges and vectors can be independently recalculated for every slow-time instant $t$ using any movement model.
    Hence, the proposed model enables the simulation of arbitrary target movements and complex trajectories, that can include varying accelerations and body rotation, which is not feasible with CAD-based EM simulations.

    For the rare scenarios where the drone moves with very high translational or rotational speeds, it is necessary to recalculate $R_O$, using {\eqref{eq:R_mu_m}}, for each fast-time sample $m$.
    In this case, the model is able to reproduce the effect of range migration of the target in a symbol duration.
    But for almost all cases, drones cannot achieve such speeds, thus it is reasonable to consider that $R_O$ is constant along the symbol duration.
    This reduces the computational complexity.

    The computational efficiency of the proposed model can even be improved by applying parallel computation using the following steps:
    firstly calculate all time position-dependent parameters; then use parallel processing to calculate each slow-time snapshot with the currently proposed model.
    This is possible due to the independence between the movement model and the proposed signature generation model.

    Additionally, the proposed model can be easily implemented, not requiring paid software frameworks or powerful servers or High-Performance Computing (HPC) systems, for generating a large amount of data. 

\section{CONCLUSION}\label{section:Conclusion}

This study proposes a model for simulating the bistatic micro-Doppler signature of multi-propeller drones in distributed ISAC. 
Micro-Doppler signature of a single propeller is derived using the thin wire model, which regards a propeller as a homogeneous, linear, rigid antenna, that can be replaced by an infinite set of point scatterers.
Then, this single propeller model is extended into a multi-propeller model.
This model allows the use of different configurations for each propeller, such as position, orientation, and rotation speed.
In addition, this model is able to generate high-range resolution (HRR) signatures.

Two methods to generate the contribution of static parts are also proposed, namely the Gaussian model and the measurement-based model.
Finally, vibration is added to the body contribution, concluding the micro-Doppler model for the multi-propeller drone.
The model is successfully validated by comparing simulated target signatures against the signatures obtained from measurements. 

The model is able to generate realistic data, comparable to those of real measurements.
Frequency domain and range-Doppler plots of data generated with the proposed model show the main effects visible in measurement data.
It can also be seen that the model reproduces the relation between the bistatic angle and the micro-Doppler spread in the same way it appears in the measurement data signature.

It is likewise concluded that including vibration in the model is vital to allow for a good representation of the drone body contribution in the high-range resolution case.
Although the measurement-based model produces a better representation of the drone body HRR micro-Doppler, it has the drawbacks of requiring multiple measurements, or at least an electromagnetic full-wave simulation, for each different target and having limitations in terms of geometric constellations.
To gain a generic representation of the target body, instead, the Gaussian body model can be applied.

Future work can comprise the use of simulated data using this proposed model to train target classification algorithms and the validation of this training method using measurement data.

{\appendices
    \section{BIRA MEASUREMENT SYSTEM AND DRONES}\label{subsection:BiRa}
    
\begin{figure}[b]
    \centering
    \includegraphics[width=0.45\textwidth, height=1.85in]{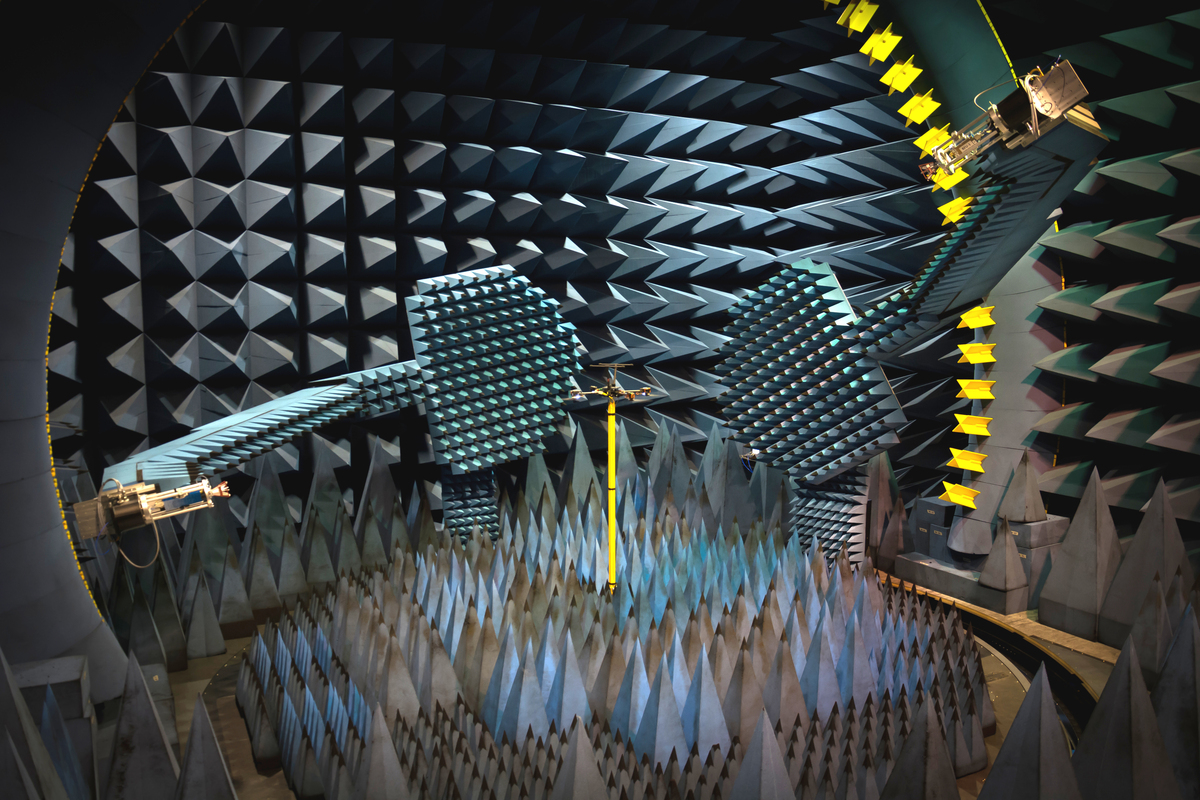}
    \caption{BIRA measurement system.}
    \label{fig:DJI}
\end{figure}

Micro-Doppler measurements, combining bistatic geometry and broadband OFDM waveform, were performed to validate the model from Sections \ref{section:Model} and \ref{section:Drone_Model}. For this task, multiple bistatic micro-Doppler measurements with different aspect angles and employing OFDM waveform were necessary.

A solution for this kind of measurement is the recently inaugurated BIRA measurement system, situated at the Thuringian Center of Innovation in Mobility (ThIMo) of the Technische Universit\"at Ilmenau. 

This advanced system comprises two pivoting gantries, equipped with a transmitting (Tx) and receiving (Rx) antenna, respectively \cite{andrich2024bira}. 
The target object of interest is positioned at the center of the turntable as in \autoref{fig:DJI}.
Therefore, the target can be illuminated and observed from any desired bistatic constellation within a coverage of \ang{360} in azimuth and, in elevation, the complete upper hemisphere, and from \ang{25} to \ang{0} in the lower hemisphere.
\begin{figure}[t]
\centering
    \includegraphics[width=3.0in]{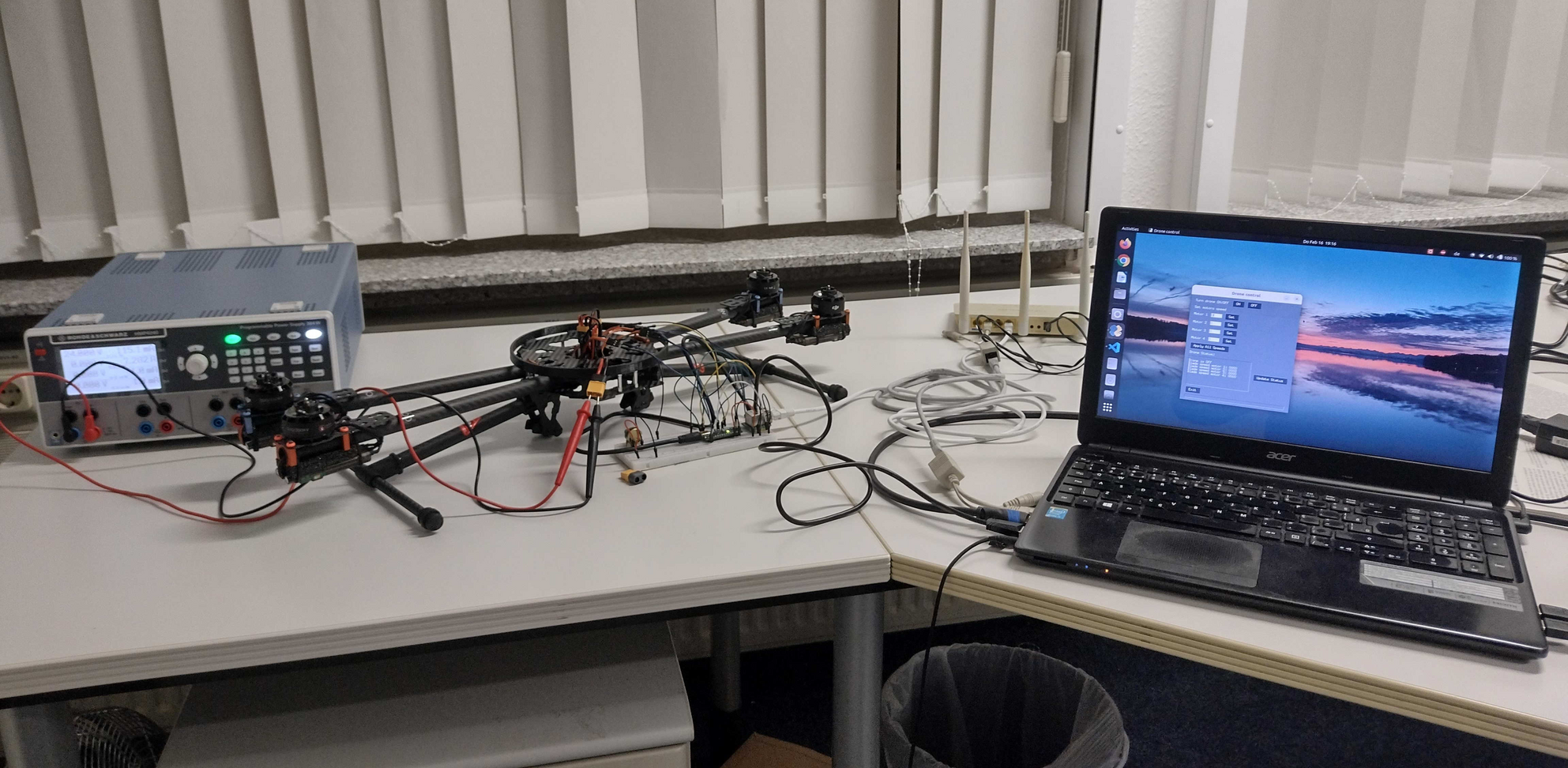}
    \caption{Custom-built target with rotation speed control for micro-Doppler measurement in BIRA.}
    \label{fig:mockup}
\end{figure}

Moreover, BIRA allows efficiently setting up sets of Tx/Rx positions, with joint control of gantries positions, turntable azimuth, and device under test (DUT) settings. 
Therefore, it is possible to perform long fully automated measurement runs, including multiple bistatic angles, without requiring any operator interaction.

A particularly noteworthy feature is that its mechanical system is modular and can be integrated with a vast variety of Tx/Rx measurement systems, thus allowing multiple combinations of applications, waveforms, bandwidths, carrier frequencies, etc. 
Equipping BIRA with the software-defined radio (SDR) system architecture detailed in \cite{rfsoc_paper} allows high range resolution (HRR) micro-Doppler measurements, with instantaneous bandwidth larger than 2 GHz.

Another important aspect of this measurement infrastructure is that for micro-Doppler analysis purposes, in order to provide reference speed, it is desirable to have full control over the rotors' rotation, which is not feasible with most standard commercial drones.
Therefore, a custom-built system was assembled, allowing us to remotely set the propeller's angular velocities. 
This system, presented on \autoref{fig:mockup}, is composed of a Tarot IRON MAN 650 mechanical structure, with motors, sensors, and an Arduino on a PCB board for setting up motors thrust values and measuring rotation speeds, as well as an Ethernet shield for remote commanding.
Carbon fiber propellers were used in the measurements.

    \section{BISTATIC REFLECTIVITY MEASUREMENT} \label{section:Static_measurements}
      
The reflectivity of an object can be described as follows for both far-field and near-field conditions \cite{Schwind, costa2024bistatic}: 
    \begin{equation} 
       S_{21}(f)=\frac{\vert E_\text{scat}(f)\vert ^{2}}{\vert E_\text{inc}(f)\vert ^{2}}.
       \label{equ_reflectivity}
    \end{equation}
Here, $E_\text{inc}$ is the electric field strength of the incident wave impinging on the target, and $E_\text{scat}$ is the electric field strength of the scattered wave.
The reflectivity of an object depends on its size, shape, and material composition, the incident and scattering wave aspects, the radar frequency, and the wave polarization. 
    
The reflectivity of Ironman Drone is measured with wide bandwidth (\SI{5.78}{GHz}:\SI{10}{MHz}:\SI{8.22}{GHz}) for multiple bistatic angle constellations (10°:5°:180°) in HH polarization.    
The measurement is performed with and without the drone.
In post-processing, the background subtraction is performed by subtracting these two measurements to eliminate the reflection from the anechoic chamber.
After that, time domain gating is applied to filter out the residual reflections.
In \autoref{fig:Reflectivity_Ironman}, the bistatic reflectivity of the Ironman Drone is plotted.
This measured reflectivity is integrated into the proposed model. 
    \begin{figure}[b]
        \centering
        \includegraphics[width=0.485\textwidth]{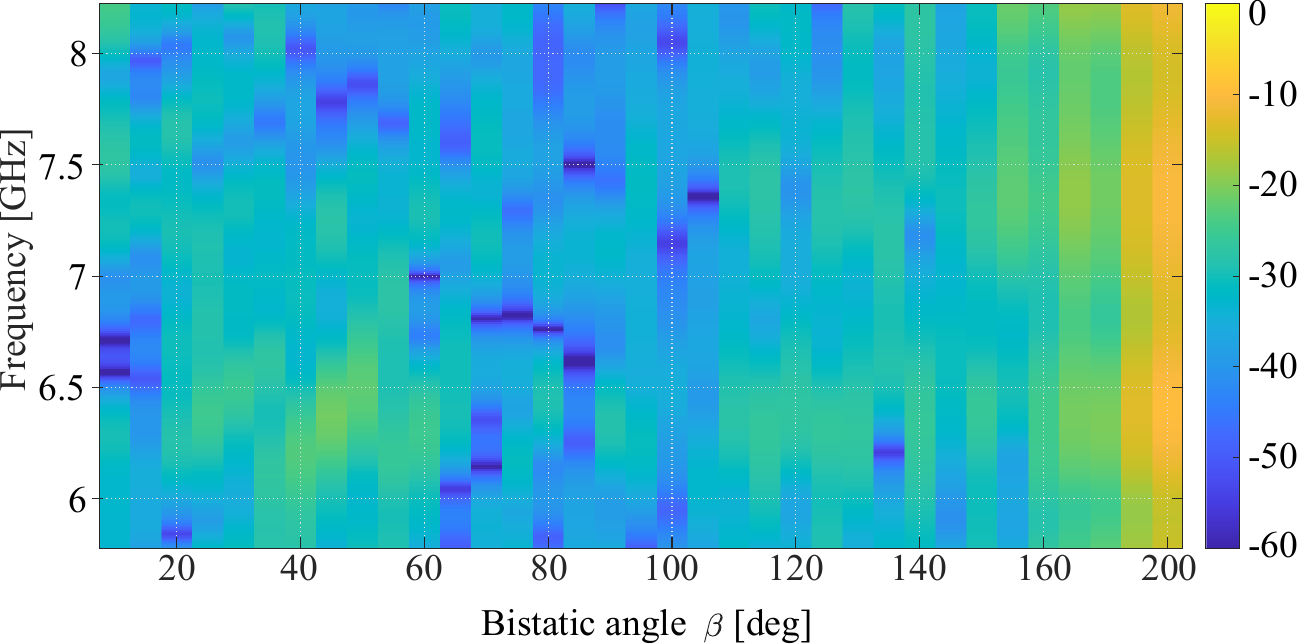}
        \caption{Measured bistatic reflectivity of Ironman drone at center frequency 7 GHz with 2.44 GHz bandwidth for bistatic angles 10° to 180°.}
        \label{fig:Reflectivity_Ironman}
    \end{figure}
      

    \section{BISTATIC MICRO-DOPPLER MEASUREMENTS} \label{subsection:BiRa_measurement}
    
\begin{table}[t]
    \caption{MICRO-DOPPLER SIMULATION AND MEASUREMENT SETUPS}
    \label{tab_MicroDoppler_measurement_setup}
    \centering
    \begin{tabular}{c l l l}
        \hline
        \textbf{Parameter} & \textbf{Setup 1} & \textbf{Setup 2} \\
        \hline
        Waveform & OFDM-based & OFDM-based \\
        \hline
        Central frequency & \SI{3.7}{GHz} & \SI{7}{GHz} \\
        \hline
        Bandwidth & \SI{200}{MHz} & \SI{2.46}{GHz} \\
        \hline
        Total number of carriers & 1600 & 2500 \\
        \hline
        Carriers with energy & 1280 & 2048 \\
        \hline
        Symbol duration & \SI{8}{\mu s} & \SI{1.02}{\mu s} \\
        \hline
        Rotating propellers & 1 & 2 \\
        \hline
        Blades per propeller & 2 & 2 \\
        \hline
        Propeller radius & \SI{16.55}{cm} & \SI{16.55}{cm} \\
        \hline
        Rotation speed & \SI{1500}{rpm} & 1500 and \SI{2000}{rpm} \\
        \hline
    \end{tabular}
\end{table}

    Two setups were implemented, using a wideband OFDM-based transmit signal called Newman sequence with constant spectral magnitude and minimal crest-factor \cite{Boyd1986}. 
    \autoref{tab_MicroDoppler_measurement_setup} details the measurement system setups.
    
    For micro-Doppler analysis, an OFDM radar processing was performed, as in \cite{braun2014ofdm}, with a back-to-back measurement as a reference for proper calibration.
    The signal was then processed in slow-time, i.e., a vector of the returns of the target in a single range bin along different symbols. 

    In order to reduce the amount of data to be processed, while focussing on a bandwidth of interest, subsampling and averaging over integer symbols can be applied, as far as this is compliant to the complex Nyquist criterion for the maximum speed.
    For example, in \autoref{fig:simul_measurement}, a subsampling factor of 8 means that we use only every eighth symbol available.

    \section{EFFECT OF INEQUALITY IN THE MODEL} \label{subsection:Inequality}
    As explained in \mbox{\autoref{subsection:geometry}}, the proposed model is derived from the geometrical description for the bistatic range of a single rotating point, developed by \mbox{\cite{Ai2011}}.

This description presents an inequality in {\eqref{eq:R_t}}, that is ignored in the following sections for simplicity.
However, for mathematical rigor, it is necessary to evaluate the impact of this inequality in the proposed model.

To achieve this evaluation, it is necessary to understand where the inequality comes from.

To start his analysis, the author considers a coordinates system (\mbox{\autoref{fig:rot_pt1}}) with origin in \textbf{O}, plane \textbf{XY} containing the rotation and axis \textbf{X} containing the projection of the transmitter over plane \textbf{XY}.
        
According to the spatial geometry, the following equations can be obtained.
\begin{equation}
    TP^2=TT'^2+T'P^2,
\end{equation}
and
\begin{equation}
    T'P^2=OT'^2+OP^2-2OT'\cdot OP \cdot .\cos(\angle T'OP).
\end{equation}

From these two equations, applying $OP=l$, $OT'=R_T\sin\beta_T$ and $TT'=R_T\cos\beta_T$, we have
\begin{equation}
    R_{PT}(t)=\sqrt{R_T^2+l^2-2R_Tl\sin\beta_T\cos(\omega t+\varphi_T)}
\end{equation}

where $R_{PT}(t)$ is the distance of the target to the transmitter, $l$ is the distance from the rotating scatterer point to the rotation center, $\omega$ is the radial speed of rotation, and $\beta_T$ and $\varphi_T$ are, respectively, the elevation and the initial azimuth aspect angles with respect to the transmitter.

If we define $C=\sin\beta_T\cos(\omega t+\varphi_T)$, we can make
\begin{equation} \label{eq:3}
    R_{PT}(t)=R_T\sqrt{1+\frac{l^2}{R_T^2}-2\frac{l}{R_T}C}
\end{equation}

Our inequality comes from the assumption of $R_T\gg l$, from which {\eqref{eq:3}} becomes
\begin{equation} \label{eq:4}
    R_{PT}(t) \approx R_T - lC.
\end{equation}

Now, we can evaluate the first order error of this approximation by using the Taylor expansion for the square root:
\begin{equation}
    \sqrt{1 + x} = 1 + \frac{1}{2}x - \frac{1}{8}x^2 + \mathcal{O}(x^3)
\end{equation}

Applying to our case, with $x=2\frac{l}{R_T} C + \frac{l^2}{R_T^2}$, we have that the error over $R_{PT} \approx R_T + lC$ is

\begin{equation}
\Delta R_{PT} = \frac{l^2}{2R_T} - \frac{R_T}{8}(2\frac{l}{R_T} C + \frac{l^2}{R_T^2})^2 + R_T\mathcal{O}(x^3)
\end{equation}

\begin{equation}
Err = \frac{l^2}{2R_T} - \frac{1}{8R_T}(4l^2 C^2 + 4l^3 C/R_T + l^4/R_T^2) + R_T\mathcal{O}(x^3)
\end{equation}

\begin{equation}
Err = \frac{l^2}{2R_T}(1 - C^2) - \frac{1}{2R_T^2}l^3C - \frac{1}{8R_T^3}l^4 + R_T\mathcal{O}(x^3)
\end{equation}

Thus, the upper bound of the error is
\begin{equation}
Err \leq \frac{l^2}{2R_T}.
\end{equation}

The calculation of the distance between target and receiver and of its respective error is analogous.
So, the upper bound of the error in the calculation of the bistatic range is given by
\begin{equation}
Err \leq \frac{l^2}{2}\left(\frac{1}{R_T}+\frac{1}{R_R}\right)
\end{equation}

In the examples used in the current paper, in order to match the measurements conditions, we used $l = \qty{0,1655}{m}$ and $R_T = R_R = \qty{3.43}{m}$.
This case results in an error of range with upper bound of $Err \leq \qty{8}{mm}$, which represents a relative error of $RErr \leq 0.12\%$ of the bistatic range. 

It is important to notice that in practical scenarios the bistatic range is usually much larger and, as a consequence, the error will be smaller.}

\section*{ACKNOWLEDGMENT}
The authors would like to thank Dr.-Ing. Tobias Nowack and M.\,Sc. Masoumeh Pourjafarian for their support throughout the BIRA measurement in ThIMo.

\bibliographystyle{IEEEtran}
\bibliography{references}
\vfill\pagebreak

\begin{IEEEbiography}[{\includegraphics[width=1in,height=1.25in,clip,keepaspectratio]{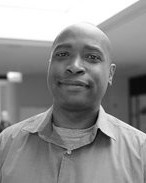}}]{Heraldo Cesar Alves Costa}~
received a B.Sc. degree in Electronic Engineering and a M.Sc. degree in Electrical Engineering from the Military Institute of Engineering (IME), Rio de Janeiro, Brazil, in 2002 and 2008, respectively.
He is currently pursuing a Ph.D. degree in Electrical Engineering at Technische Universität Ilmenau, Germany. His research interests include micro-Doppler signatures, modeling bistatic dynamic reflectivity of objects, and multistatic target classification in Integrated Communication and Sensing (ICAS) applications.
\end{IEEEbiography}
\vspace{-5 mm}
\begin{IEEEbiography}[{\includegraphics[width=1in,height=1.25in,clip,keepaspectratio]{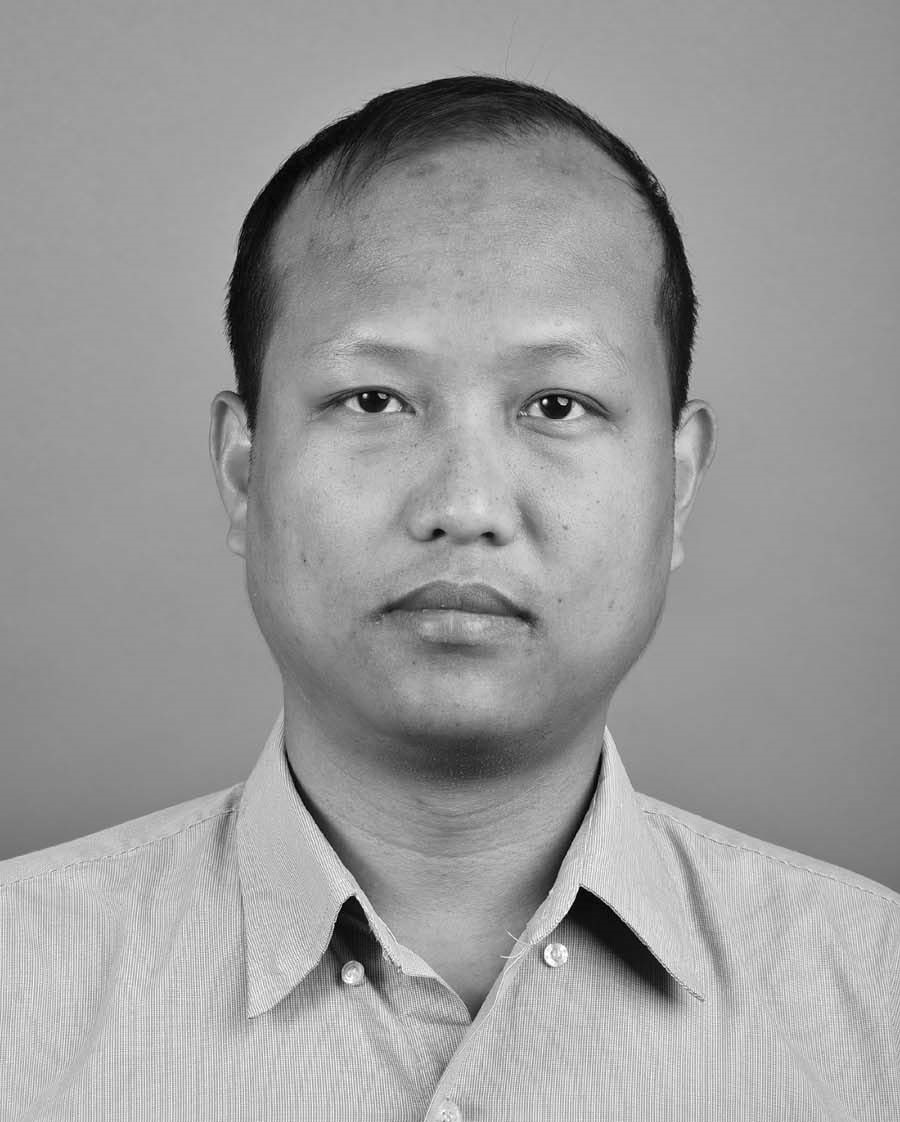}}]{Saw James Myint}~
received a B.E. degree in Electronic Engineering from Mandalay Technological University, Myanmar, in 2005, an M.E. degree in Telecommunication Systems from Moscow Aviation Institute, Russian Federation, in 2011, and an M.Sc. degree in Communications and Signal Processing from Technische Universität Ilmenau, Germany in 2018. He is currently pursuing a Ph.D. degree in Electrical Engineering at Technische Universität Ilmenau, Germany. His research interests include multistic radar, modeling multistatic reflectivity/RCS of objects, and target classification in Integrated Communication and Sensing (ICAS) applications.

\end{IEEEbiography}
\vspace{-5 mm}
\begin{IEEEbiography}[{\includegraphics[width=1in,height=1.25in,clip,keepaspectratio]{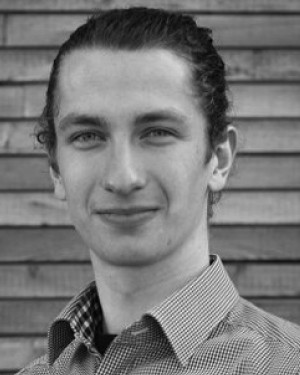}}]{Sebastian W. Giehl}~
received the B.Sc. and M.Sc. degrees in electrical engineering and information technology from Technische Universität Ilmenau, Germany, in 2020 and 2022, respectively. He has been working on parallel computing FPGA systems, since 2017. In 2022, he was with Fraunhofer IIS, Germany. Since 2023, he has been a Research Assistant with Technische Universität Ilmenau on ISAC in mobility applications for the next generation of mobile communication and FPGA-based, high-bandwidth, and real-time capable SDR systems, for e.g., radar or channel sounding.
\end{IEEEbiography}
\vspace{-5 mm}
\begin{IEEEbiography}[{\includegraphics[width=1in,height=1.25in,clip,keepaspectratio]{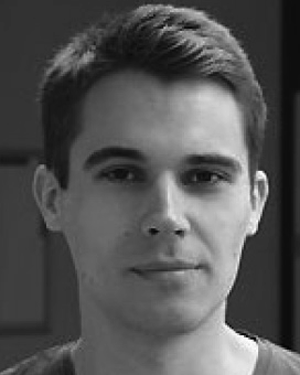}}]{Carsten Andrich}~
received the B.Sc. and M.Sc. degrees in electrical engineering and information technology from Technische Universität Ilmenau, Ilmenau, Germany, in 2014 and 2016, respectively. He is currently pursuing a Dr.-Ing. degree. He is a Researcher with the Fraunhofer Institute for Integrated Circuits IIS, Ilmenau. His current research interests include high-precision measurement applications for software-defined radio devices and efficient digital signal processing algorithm development.
\end{IEEEbiography}
\vspace{-5 mm}
\begin{IEEEbiography}[{\includegraphics[width=1in,height=1.25in,clip,keepaspectratio]{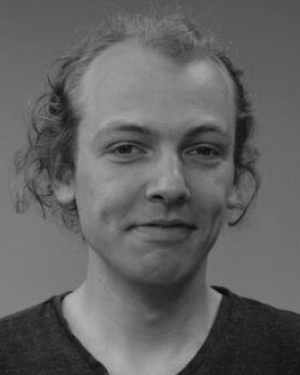}}]{Maximilian Engelhardt}~
received the Bachelor of Science (B.Sc.) and Master of Science (M.Sc.) degrees in computer and systems engineering from Technische Universität Ilmenau, Germany, in 2019 and 2021, respectively. Since 2021, he has been a Research Assistant with Fraunhofer IIS, Germany. His research interests include novel software-define radio solutions and radio-frequency measurement architectures, whereby his focus is on implementing real-time systems on off-the-self hardware.
\end{IEEEbiography}
\vspace{-5 mm}
\begin{IEEEbiography}[{\includegraphics[width=1in,height=1.25in,clip,keepaspectratio]{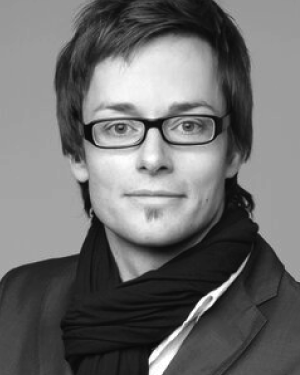}}]{Christian Schneider}~
received the Diploma degree in electrical engineering from Technische Universität Ilmenau, Germany, in 2001. He is currently the Group Leader of the Electronic Measurements and Signal Processing Department (EMS), Technische Universität Ilmenau and Fraunhofer IIS. His research interests include multi-dimensional channel sounding, radio channel characterization and modeling, and its application to space-time signal processing and ISAC questions. He received the Best Paper Award at the European Wireless Conference in 2013 and the European Conference of Antennas and Propagation in 2017 and 2019.
\end{IEEEbiography}
\vspace{-5 mm}
\begin{IEEEbiography}[{\includegraphics[width=1in,height=1.25in,clip,keepaspectratio]{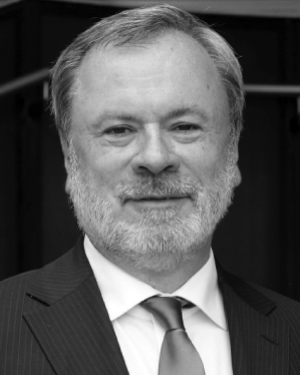}}]{Reiner S. Thom\"a}~
received a degree in electrical engineering and information technology from TU Ilmenau, Germany. Since 1992, he has been a Professor at TU Ilmenau. He has retired, since 2018. In 2007, he received the Thuringian State Research Award for Applied Research and the Vodafone Innovation Award, in 2014, both for his contributions to high-resolution multidimensional channel sounding. In 2020, he received the EurAAP Propagation Award “For pioneering the multi-dimensional description of the mobile radio channel by advanced signal-processing methods.” He has contributed to several European and German research projects and clusters. His research interests include multidimensional channel sounding, propagation measurement and model-based parameter estimation, MIMO system over-the-air testing in virtual electromagnetic environments, MIMO radar, passive coherent location, and integrated sensing and communication.
\end{IEEEbiography}

\vfill\pagebreak

\end{document}